\documentclass
[aps,prb,eqsecnum,showpacs,superscriptaddress,floats,amsmath]{revtex4}%
\usepackage{graphicx}
\usepackage{amsmath}
\usepackage{amsfonts}
\usepackage{amssymb}%
\newcommand{\be}{\begin{equation}}
\newcommand{\ee}{\end{equation}}

\begin{document}
\author{Fernando~M.~Cucchietti}
\affiliation{ {\it Theoretical Division, MS B213, Los Alamos National
Laboratory,}
Los Alamos, NM 87545}
\affiliation{ {\it Facultad de Matem\'{a}tica, Astronom\'{\i}a y
F\'{\i}sica,Universidad Nacional de C\'{o}rdoba,} \\
Ciudad Universitaria, 5000 C\'{o}rdoba, Argentina}
\author{Horacio~M.~Pastawski}
\affiliation{ {\it Facultad de Matem\'{a}tica, Astronom\'{\i}a y
F\'{\i}sica,Universidad Nacional de C\'{o}rdoba,} \\
Ciudad Universitaria, 5000 C\'{o}rdoba, Argentina}
\author{Rodolfo~A.~Jalabert}
\affiliation{ {\it Institut de Physique et Chimie des
Mat\'{e}riaux de Strasbourg, UMR 7504, CNRS-ULP,}\\
{\it 23 rue du Loess, BP 43, 67034 Strasbourg Cedex 2, France} }
\title{Universality of the Lyapunov regime for the Loschmidt echo}
\date{July 30, 2003}

\begin{abstract}
The Loschmidt echo (LE) is a magnitude that measures the sensitivity of
quantum dynamics to perturbations in the Hamiltonian. For a certain regime of
the parameters, the LE decays exponentially with a rate given by the Lyapunov
exponent of the underlying classically chaotic system. We develop a
semiclassical theory, supported by numerical results in a Lorentz gas model,
which allows us to establish and characterize the universality of this
Lyapunov regime. In particular, the universality is evidenced by the
semiclassical limit of the Fermi wavelength going to zero, the behavior for
times longer than Ehrenfest time, the insensitivity with respect to the form
of the perturbation and the behavior of individual (non-averaged) initial
conditions. Finally, by elaborating a semiclassical approximation to the
Wigner function, we are able to distinguish between classical and quantum
origin for the different terms of the LE. This approach renders an
understanding for the persistence of the Lyapunov regime after the Ehrenfest
time, as well as a reinterpretation of our results in terms of the
quantum--classical transition.

\end{abstract}
\maketitle

\section{\bigskip Introduction}

\label{sec:intro}

Controlling the phase in the evolution of a quantum system is a fundamental
problem that is becoming increasingly relevant in many areas of physics. In
relatively simple systems, like a quantum dot in an Aharanov-Bohm
ring\cite{cit-HeiblumNature}, the phase can even be measured by transport
experiments. The development of the quantum information field requires the
control of the phase of increasingly complex
systems\cite{cit-DiVincenzoNature}. Such a control is hindered by interactions
with the environment in a way which is not completely understood at present.

Nuclear Magnetic Resonance provides a privileged framework to test our ideas
on the evolution and degradation of the quantum phase. The phenomenon of spin
echo, through the reversal of the time evolution, allows to study how an
individual spin, in an ensemble, loses its phase memory\cite{cit-HannEcho}.
The randomization of its phase appears as a consequence of the interaction
with other spins that act as an environment. Recently, it has become possible
to test the phase of the collective many-spin state through the experiments of
Magic\cite{cit-Pines} and Polarization\cite{cit-Polecho} echoes. In these
cases a local polarization \textquotedblleft diffuses" away as consequence of
the spin-spin interactions in the effective Hamiltonian $\mathcal{H}$. The
whole many-body dynamics is then reversed by the sudden transformation
$\mathcal{H}\rightarrow-\mathcal{H}$ . However, there is an increasing failure
to reach the initial polarization state which is a consequence of the
fluctuations of the phase of the complex quantum state\cite{cit-JChemPhys} and
is a measure of the entropy growth\cite{cit-MolPhysics}.

Surprisingly, the rate of loss of information of the phase appears as an
intrinsic property of the system, being quite insensitive to how small is the
coupling to the external degrees of freedom or the precision of the
reversal\cite{cit-Potencia}. This may be interpreted as analogous to the
residual resistivity of impure metals. When the direct coupling to the thermal
bath is decreased by lowering the temperature, the resistivity becomes
controlled by the reversible elastic scattering with impurities
\cite{cit-Zeeman,cit-Laughlin}. The common feature of both intrinsic behaviors
is the complexity of the dynamics that justifies the \textit{stosszahlansatz}
or molecular chaos hypothesis.

However, such an hypothesis does not seem to be compatible with our basic
knowledge of quantum dynamics. Unlike classical mechanics, quantum dynamics
exhibits a remarkable insensitivity to initial conditions
\cite{cit-QChaos-europ,cit-Haake}. That is why the field known as quantum
chaos deals mainly with the quantum stationary properties of systems whose
underlying classical dynamics is chaotic. Among these properties, the ones
more frequently studied are the level statistics\cite{cit-Bohigas}, wave
function scarring \cite{cit-scars}, and parametric correlations
\cite{cit-Aaron}. A notable exception among these studies was that of Peres
\cite{cit-Peres}, who realized that classically chaotic and integrable systems
behave differently under imperfect time reversal, for very short and long
times. It is through the experimental findings above cited, that the study of
time evolution of classically chaotic systems has gained a privileged place in
nowadays research.

A simplified version of the echoes experimentally studied is the so-called
Loschmidt echo (LE)%

\begin{equation}
M(t)=\left\vert m(t)\right\vert ^{2}=\left\vert \left\langle \psi
_{0}\right\vert e^{\mathrm{i}(\mathcal{H}_{0}+\Sigma)t/\hbar}\ e^{-\mathrm{i}%
\mathcal{H}_{0}t/\hbar}\left\vert \psi_{0}\right\rangle \right\vert ^{2},
\label{eq-LoschmidtEcho}%
\end{equation}
where $\left\vert \psi_{0}\right\rangle $ is an arbitrary initial state that
evolves forward in time under the system Hamiltonian $\mathcal{H}_{0}$ for a
time $t$, and then backwards under a slightly perturbed Hamiltonian
$\mathcal{H}_{0}+\Sigma$. The amplitude $m(t)$ of the LE is the overlap
between the two slightly different evolutions of the same initial state, and
$M(t)$ quantifies the departure from the perfect overlap. Because of this
important property, within the field of Quantum Information the LE is referred
to as \textquotedblleft fidelity" \cite{Chuang}. Alternatively, $M(t)$ can
also be written as the trace of the product of two pure-state density matrices
$\rho$ or Wigner functions $W$ evolving with different Hamiltonians,%

\begin{equation}
M(t)=\mathop{\rm tr}\{\rho_{\mathcal{H}_{0}+\Sigma}(t)\ \rho_{\mathcal{H}_{0}%
}(t)\}=(2\pi\hbar)^{d}\int\mathrm{d}\mathbf{r}\int\mathrm{d}\mathbf{p}%
\ W_{\mathcal{H}_{0}+\Sigma}(\mathbf{r},\mathbf{p};t)\ W_{\mathcal{H}_{0}%
}(\mathbf{r},\mathbf{p};t)\ . \label{eq-LEWigner}%
\end{equation}
We have used the standard definitions%

\begin{equation}
\rho_{\mathcal{H}} = \left|  \psi\right\rangle \left\langle \psi\right|  \ ,
\ \text{with} \ \left|  \psi\right\rangle = e^{-\mathrm{i}\mathcal{H}t/\hbar
}\left|  \psi_{0}\right\rangle , \label{eq:DEFRW0}%
\end{equation}

\begin{equation}
W_{\mathcal{H}}(\mathbf{r},\mathbf{p};t)=\frac{1}{(2\pi\hbar)^{d}}%
\int\mathrm{d}\delta\mathbf{r}\ \exp{\left[  -\frac{\mathrm{i}}{\hbar
}\ \mathbf{p}\cdot\delta\mathbf{r}\right]  }\ \left\langle \mathbf{r}%
+\frac{\delta\mathbf{r}}{2}\right\vert \rho_{\mathcal{H}}\left\vert
\mathbf{r}-\frac{\delta\mathbf{r}}{2}\right\rangle , \label{eq:DEFRW1}%
\end{equation}
where $d$ is the dimensionality of the space.

In consistency with the experimental behavior of the polarization echo, the LE
of a classically chaotic one-body Hamiltonian $\mathcal{H}_{0}$ was found to
exhibit an intrinsic decay rate \cite{cit-Jalabert-Past}. This result is valid
beyond some critical value of the perturbation. Interestingly, the decay rate
is precisely the Lyapunov exponent $\lambda$ of the classical system. A
related relevance of the classical dynamics had been hinted from the analysis
of the entropy growth of dissipative systems \cite{cit-Zurek-Paz}.

The purely Hamiltonian character of the model of
Ref.~\onlinecite{cit-Jalabert-Past}, as well as the result of a classical
parameter ($\lambda$) governing a bona fide quantum property ($M$), attracted
considerable attention. A quite intense activity has been devoted in the last
two years in order to test these predictions in various model systems and
pursue further developments of the theory \cite{cit-PhysicaA}$^{-}%
$\cite{cit-Mirlin}.

The Lyapunov behavior has been numerically obtained in models of a Lorentz gas
\cite{cit-Lorentzgas}, kicked tops\cite{cit-Jacquod01,cit-WangLi}, Bunimovich
stadium \cite{cit-Bunimovich}, bath tube stadium \cite{cit-SmoothBilliard} and
sawtooth map \cite{cit-BenentiCasati}. The analytical results have been mainly
focused in the small perturbation region. Jacquod and collaborators
\cite{cit-Jacquod01} identified the regime below the critical perturbation as
following a Fermi Golden Rule through the energy uncertainty produced by the
perturbation which were also analyzed with semiclassical tools
\cite{cit-CerrutiTomsovic}. Prosen and collaborators
\cite{cit-Prosen,cit-ProsenSeligman} showed that $M(t)$ in the perturbative
regime depends on the specific time dependence of the perturbation correlation functions.

The Lyapunov regime bears a clear signature of the underlying classical
dynamics. This observation lead Benenti and Casati \cite{cit-BenentiCasati} to
propose that the independence of the decay rate on the perturbation strength
is a consequence of the quantum-classical correspondence principle. As we will
analyze in this work, the situation is far less trivial. We are going to show
that this regime persists for times much larger than the Ehrenfest time (as
defined by Berman and Zavslavsky \cite{cit-BermanZavslasky}). In addition, the
quantum LE is functionally different than what a direct estimation would yield
for the classical LE (for the chaotic \cite{cit-Jalabert-Past}, as well as the
integrable \cite{cit-JacquodInt} cases). Moreover, the classical counterpart
of the LE is problematic since a wide range of dynamic behaviors is obtained
in different situations \cite{cit-Eckart,cit-CasatiClassic}.

The LE has also been studied in different disordered systems
\cite{cit-PhysicaA,cit-Mirlin}. It has been shown in both cases that the long
range of the perturbing potential, as emphasized in Ref. \onlinecite
{cit-Jalabert-Past}, is crucial in order to obtain a perturbation independent regime.

The various approximations that the semiclassical theory of
Ref.~\onlinecite{cit-Jalabert-Past} relies on were further corroborated using
an initial momentum representation of the wave--packet \cite{Heller}. This
changes the sum over an uncontrolled number of trajectories into only one,
which allows the exact numerical evaluation of the semiclassical expression
for the echo.

Taking the perturbation as the action of an external environment allows us to
think of the LE as a measure of the decoherence. This approach has been
advocated by Zurek\cite{cit-ZurekNature}, and extended\cite{cit-LESubPlanck}
by studying the decay of $M(t)$ as expressed by a product of Wigner functions
(Eq. (\ref{eq-LEWigner})). A semiclassical approximation to the Wigner function
allows us to separate the different
contributions to the LE coming from classical and non-classical processes. As
we discuss in detail in the sequel, such distinction enables to quantify how
decoherence builds in until the classical terms finally dominate the LE.

With the goal of addressing experimentally relevant systems
\cite{cit-Weiss,cit-Richter,cit-Portal}, we illustrate our findings in a
simple model with classical chaotic dynamics: the Lorentz gas. This system has
been shown to exhibit a well defined Lyapunov regime \cite{cit-Lorentzgas}.
The semiclassical theory that we develop, as well as the extensive numerical
results that we present in this work, allows us to establish and characterize
the universality of the perturbation independent regime.

This universality manifests itself by the robustness of the Lyapunov regime
with respect to various effects. Firstly, in the semiclassical limit of Fermi
wavelength $\lambda_{F}$ going to zero, the borders of the regime extend from
zero perturbation up to a classical upper bound. Secondly, and as stated
above, for finite $\lambda_{F}$ the Lyapunov regime extends up to times
arbitrarily larger than Ehrenfest's time. Finally, universality is also
evidenced by the insensitivity of the Lyapunov regime with respect to the form
of the perturbation or the (non-averaged) behavior of individual semiclassical
initial conditions.

The paper is organized as follows: in section~\ref{sec:semi} we develop the
semiclassical approach to the LE with a quenched disorder playing the role of
the perturbation, as proposed in Ref.~\onlinecite {cit-Jalabert-Past}. We then
discuss the main assumptions and set the theoretical framework that will be
further developed in the rest of the paper. In section~\ref{sec:lorentz} we
consider a specific model, the Lorentz gas, and a different perturbation than
in the previous case. We first characterize the classical dynamics of the
Lorentz gas, as well as that of the perturbation, and then present a
semiclassical calculation of the LE, discussing the different regimes
predicted by the theory. In section~\ref{sec:universality} we concentrate the
main results of this work. The universality of the Lyapunov regime is
discussed and supported with numerical results on the semiclassical limit, the
behavior after the Ehrenfest time and the effects of averaging. In
section~\ref{sec:wigner} we discuss the relation of the LE to decoherence by
studying the semiclassical approximation to the Wigner function and
reinterpreting the results of Sec.~\ref{sec:semi} under this new highlight. We
conclude in section~\ref{sec:conclusions} with some final remarks.

\section{The Loschmidt echo - Semiclassical analysis}

\label{sec:semi}

\subsection{Semiclassical evolution}

\label{subsec:semievo}

In this section we calculate the Loschmidt echo (Eq. (\ref{eq-LoschmidtEcho}))
for a generic chaotic system $\mathcal{H}_{0}$ and a perturbation $\Sigma$
arising from a quenched disorder. We follow the analytical scheme of
Ref.~\onlinecite{cit-Jalabert-Past}, discussing the main assumptions and the
generality of the results. We choose as initial state a Gaussian wave-packet
(of width $\sigma$), which is the closest we can get to a classical state.%

\begin{equation}
\psi(\overline{\mathbf{r}},t\!=\!0)=\left(  \frac{1}{\pi\sigma^{2}}\right)
^{d/4}\exp\left[  \frac{\mathrm{i}}{\hbar}\mathbf{p}_{0}\cdot(\overline
{\mathbf{r}}-\mathbf{r}_{0})-\frac{1}{2\sigma^{2}}(\overline{\mathbf{r}%
}-\mathbf{r}_{0})^{2}\right]  . \label{eq-PaqueteInicial}%
\end{equation}

\noindent We will keep the spatial dimension $d$ arbitrary in the analytical
calculations, but it will be fixed to $d=2$ for the numerical studies of the
Sec.~\ref{sec:universality}. It has been shown \cite{cit-JacquodSP}
that if the initial state is a
superposition of $N$ Gaussians, the final result is the same exponential decay
one obtains with a single Gaussian but normalized by $N$. Thus, the assumption
of Eq. (\ref{eq-PaqueteInicial}) is as general as the decomposition of a given
initial state into a sum of Gaussians. The time evolution of state
$\psi(\overline{\mathbf{r}},0)$ is given by%

\begin{equation}
\psi(\mathbf{r},t)=\int\mathrm{d}\overline{\mathbf{r}}\ K(\mathbf{r}%
,\overline{\mathbf{r}};t)\ \psi(\overline{\mathbf{r}},0)\ ,
\label{eq:defprop}
\end{equation}
with the propagator
\begin{equation}
K(\mathbf{r,}\overline{\mathbf{r}};t)=\left\langle \mathbf{r}\right\vert
e^{-\mathrm{i}\mathcal{H}t/\hbar}\left\vert \overline{\mathbf{r}}\right\rangle
.
\end{equation}

We will use the semiclassical expansion of the propagator
\cite{cit-Gutz,bookBrack} as a sum over classical trajectories $s(\overline
{\mathbf{r}},\mathbf{r},t),$ going from $\overline{\mathbf{r}}$ to
$\mathbf{r}$ in a time $t,$
\begin{align}
K(\mathbf{r,}\overline{\mathbf{r}};t)  &  =\sum_{s(\overline{\mathbf{r}%
},\mathbf{r},t)}K_{s}(\mathbf{r,}\overline{\mathbf{r}};t) \ ,\nonumber\\
K_{s}(\mathbf{r,}\overline{\mathbf{r}};t)  &  =\left(  \frac{1}{2\pi i\hbar
}\right)  ^{d/2}C_{s}^{1/2} \ \exp{\left[  \frac{\mathrm{i}}{\hbar}%
S_{s}(\mathbf{r},\overline{\mathbf{r}};t) -\mathrm{i}\frac{\pi}{2}\mu
_{s}\right]  } \ , \label{eq-SemiclasicalPropagator}%
\end{align}
valid in the limit of large energies for which the de Broglie wavelength
$(\lambda_{F}=2\pi/k_{F}=2\pi\hbar/p_{0})$ is the minimal length scale.
$S_{s}(\mathbf{r,}\overline{\mathbf{r}};t)=\int_{0}^{t}d\overline{t}%
L_{s}(q_{s}(\overline{t}),\dot{q}_{s}(\overline{t});\overline{t})$ is the
action over the trajectory $s,$ and $L$ the Lagrangian. The Jacobian
$C_{s}=\left|  \det B_{s}\right|  $ accounts for the conservation of classical
probabilities, with the matrix%

\begin{equation}
\left(  B_{s}\right)  _{ij}=-\frac{\partial^{2}S_{s}}{\partial\mathbf{r}%
_{i}\partial\overline{\mathbf{r}}_{j}} \ , \label{eq:matB}%
\end{equation}

\noindent obtained from the derivatives of the action respect to the various
components of the initial and final positions. We note $\mu_{s}$ the Maslov
index, counting the number of conjugate points of the trajectory $s.$ Since we
will work with fairly concentrated initial wave-packets, we use that $\left.
\nabla_{\overline{\mathbf{r}}_{\mathrm{i}}}S_{s}\right|  _{\overline
{\mathbf{r}}=\mathbf{r}_{0}} = -{\overline{\mathbf{p}}}_{s,\mathrm{i}}$
(${\overline{\mathbf{p}}}_{s,\mathrm{i}}$ is the i-th component of the initial
momentum of the trajectory $s$) and we expand the action as%

\begin{equation}
S_{s}(\mathbf{r},\overline{\mathbf{r}};t)\simeq S_{\hat s}(\mathbf{r}%
,\mathbf{r}_{0};t)- {\overline{\mathbf{p}}}_{\hat s}\cdot(\overline
{\mathbf{r}}-\mathbf{r}_{0}) \ . \label{eq-ActionExpansion}%
\end{equation}

\noindent We are lead to work with trajectories ${\hat s}$ that join
$\mathbf{r}_{0}$ to $\mathbf{r}$ in a time $t$, which are slightly modified
with respect to the original trajectories $s(\overline{\mathbf{r}}%
,\mathbf{r},t)$. We can therefore write%

\begin{align}
\psi(\mathbf{r},t)  &  =\sum_{s(\mathbf{r}_{0},\mathbf{r},t)}K_{s}%
(\mathbf{r,r}_{0};t)\int\mathrm{d}\overline{\mathbf{r}}\ \exp{\left[
-\frac{\mathrm{i}}{\hbar}{\overline{\mathbf{p}}}_{s}\cdot(\overline
{\mathbf{r}}-\mathbf{r}_{0})\right]  }\ \psi(\overline{\mathbf{r}%
},0)\nonumber\\
&  =\left(  4\pi\sigma^{2}\right)  ^{d/4}\sum_{s(\mathbf{r}_{0},\mathbf{r}%
,t)}K_{s}(\mathbf{r},\mathbf{r}_{0};t)\ \exp{\left[  -\frac{\sigma^{2}}%
{2\hbar^{2}}\left(  {\overline{\mathbf{p}}}_{s}-\mathbf{p}_{0}\right)
^{2}\right]  }\ , \label{eq-PaqueteSemiclasico}%
\end{align}

\noindent where we have neglected second order terms of $S$ in $(\overline
{\mathbf{r}}-\mathbf{r}_{0})$ since we assume that the initial wave packet is
much larger than the Fermi wavelength ($\sigma\gg\lambda_{F}$).
Eq.~(\ref{eq-PaqueteSemiclasico}) shows that only trajectories with initial
momentum ${\overline{\mathbf{p}}}_{s}$ closer than $\hbar/\sigma$ to
$\mathbf{p}_{0}$ are relevant for the propagation of the wave-packet.

\subsection{Semiclassical Loschmidt echo}

The amplitude of the Loschmidt echo, defined in Eq.~(\ref{eq-LoschmidtEcho}),
for the initial condition (\ref{eq-PaqueteInicial}), can be approximated
semiclassically as%

\begin{equation}
m(t)=\left(  \frac{\sigma^{2}}{\pi\hbar^{2}}\right)  ^{d/2}\int\mathrm{d}%
\mathbf{r}\sum_{s,\tilde{s}}C_{s}^{1/2}C_{\tilde{s}}^{1/2}\ \exp{\left[
\frac{\mathrm{i}}{\hbar}(S_{s}-S_{\tilde{s}})-\frac{\mathrm{i}\pi}{2}(\mu
_{s}-\mu_{\tilde{s}})\right]  }\ \exp{\left[  -\frac{\sigma^{2}}{2\hbar^{2}%
}\left(  \left(  {\overline{\mathbf{p}}}_{s}-\mathbf{p}_{0}\right)
^{2}+\left(  {\overline{\mathbf{p}}}_{\tilde{s}}-\mathbf{p}_{0}\right)
^{2}\right)  \right]  }\ . \label{eq-EchoAmplitude}%
\end{equation}

Without perturbation ($\Sigma=0$) and restricting ourselves to the terms with
$s=\tilde{s}$ (which leaves aside terms with a highly oscillating phase) we
simply have%

\begin{equation}
m(t)=\left(  \frac{\sigma^{2}}{\pi\hbar^{2}}\right)  ^{d/2}\int\mathrm{d}%
\mathbf{r}\sum_{s(\mathbf{r}_{0},\mathbf{r},t)}C_{s}\ \exp{\left[
-\frac{\sigma^{2}}{\hbar^{2}}\left(  {\overline{\mathbf{p}}}_{s}%
-\mathbf{p}_{0}^{2}\right)  ^{2}\right]  }=1\ . \label{eq:mwse0}%
\end{equation}

\noindent We have performed the change from the final position variable
$\mathbf{r}$ to the initial momentum ${\overline{\mathbf{p}}}_{s}$ using the
Jacobian $C$, and then carried out a simple Gaussian integration over the
variable ${\overline{\mathbf{p}}}_{s}.$

For perturbations $\Sigma$ that are classically weak (as not to change
appreciably the trajectories governed by the dynamics of $\mathcal{H}_{0}$),
we can also neglect the terms of (\ref{eq-EchoAmplitude}) with $s\neq\tilde
{s}$ and write%

\begin{equation}
m(t)\simeq\left(  \frac{\sigma^{2}}{\pi\hbar^{2}}\right)  ^{d/2}\int
\mathrm{d}\mathbf{r}\ \sum_{s}\ C_{s}\ \exp\left[  \frac{\mathrm{i}}{\hbar
}\Delta S_{s}\right]  \ \exp{\left[  -\frac{\sigma^{2}}{\hbar^{2}}\left[
\left(  {\overline{\mathbf{p}}}_{s}-\mathbf{p}_{0}\right)  ^{2}\right]
\right]  }\ . \label{eq-mAmplSemiclassico}%
\end{equation}

\noindent Where $\Delta S_{s}$ is the modification of the action, associated
with the trajectory $s$, by the effect of the perturbation $\Sigma$. It can be
obtained as%

\begin{equation}
\Delta S_{s}= - \int_{0}^{t}d\overline{t}\ \Sigma_{s}(\mathbf{q}(\overline
{t}),\mathbf{\dot{q}(}\overline{t})) \ , \label{eq:DeltaS}%
\end{equation}

\noindent in the case where the perturbation appears as a potential energy in
the Hamiltonian (like we discuss in this chapter). If the perturbation is in
the kinetic term of the Hamiltonian (like in Sec.~\ref{sec:lorentz}), there is
an irrelevant change of sign.

Clearly individual classical trajectories will be exponentially sensitive to
perturbations and the diagonal approximation of
Eq.~(\ref{eq-mAmplSemiclassico}) would sustain only for logarithmically short
times. However, it has been argued \cite{Heller} that this approximation is
valid for much longer times because of the structural stability of the
manifold \cite{cit-CerrutiTomsovic} which allows for the existence of
trajectories arriving at $\mathbf{r}$ and departing exponentially close to
$\mathbf{r}_{0}$.

Within the approximation of Eq.~(\ref{eq-mAmplSemiclassico}), the LE is
expressed as a double integral containing two trajectories,
\begin{equation}
M(t)=\left(  \frac{\sigma^{2}}{\pi\hbar^{2}}\right)  ^{d}\int\mathrm{d}%
\mathbf{r}\int\mathrm{d}\mathbf{r}^{\prime}\sum_{s(\mathbf{r}_{0}%
,\mathbf{r},t)}\sum_{s^{\prime}(\mathbf{r}_{0},\mathbf{r}^{\prime},t)}%
C_{s}C_{s^{\prime}}\exp\left[  \frac{\mathrm{i}}{\hbar}\left(  \Delta
S_{s}-\Delta S_{s^{\prime}}\right)  \right]  \exp\left[  -\frac{\sigma^{2}%
}{\hbar^{2}}\left[  \left(  \mathbf{p}_{s}-\mathbf{p}_{0}\right)  ^{2}+\left(
\mathbf{p}_{s^{\prime}}-\mathbf{p}_{0}\right)  ^{2}\right]  \right]  .
\label{eq-MCompleto}%
\end{equation}

As in Ref.~\onlinecite{cit-Jalabert-Past}, we can decompose the LE as%

\begin{equation}
M(t)=M^{\mathrm{nd}}(t)+M^{\mathrm{d}}(t) \ , \label{eq:decomposition}%
\end{equation}

\noindent where the first term (non-diagonal) contains trajectories $s$ and
${s}^{\prime}$ exploring different regions of phase space, while in the second
(diagonal) ${s}^{\prime}$ remains close to $s$. Such a distinction is
essential when considering the effect of the perturbation over the different contributions.

\subsection{Quenched disorder as a perturbation}

In order to calculate the different components to the LE
(Eqs.~(\ref{eq-MCompleto}) and (\ref{eq:decomposition})) we need to
characterize the perturbation $\Sigma$. One possible choice
\cite{cit-Jalabert-Past} is a quenched disorder given by $N_{i}$ impurities
with a Gaussian potential characterized by the correlation length $\xi$,%

\begin{equation}
\Sigma=\tilde{V}(\mathbf{r})=\sum_{\alpha=1}^{N_{i}} \ \frac{u_{\alpha}}%
{(2\pi\xi^{2})^{d/2}}\ \ \exp{\left[  -\frac{1}{2\xi^{2}}\left(
\mathbf{r}\!-\!\mathbf{R}_{\alpha}\right)  ^{2} \right]  } \ .
\label{eq:irrevpart}%
\end{equation}

\noindent The independent impurities are uniformly distributed (at positions
$\mathbf{R}_{\alpha}$) with density $n_{i} =N_{i}/\mathtt{V}$, ($\mathtt{V}$
is the sample volume). The strengths $u_{\alpha}$ obey $\langle u_{\alpha
}u_{\beta}\rangle=u^{2}\delta_{\alpha\beta}$. The correlation function of the
above potential is given by%

\begin{equation}
C_{\tilde{V}}(|\mathbf{q}-\mathbf{q}^{\prime}|) = \langle\tilde{V}%
(\mathbf{q})\tilde{V}(\mathbf{q}^{\prime})\rangle= \frac{u^{2}n_{i}}{(4\pi
\xi^{2})^{d/2}}\ \ \exp{\left[  -\frac{1}{4\xi^{2}}(\mathbf{q}-\mathbf{q}%
^{\prime})^{2}\right]  } \ . \label{eq:corr}%
\end{equation}

The perturbation (\ref{eq:irrevpart}) does not lead to the well-known physics
of disordered systems, since the potential $\tilde{V}$ is not part of
$\mathcal{H}_{0}$, but of $\Sigma$. Then, it acts only in the backwards
propagation of the LE setup. On the other hand, the analogy with standard
disordered systems is very useful for the analytical developments. The finite
range of the potential allows to apply the semiclassical tool (provided $\xi
k_{F}\gg1$), as has been extensively used in the calculation of the orbital
response of weak disordered quantum dots \cite{RapComm,JMP,tstimp,Kbook}. The
finite range of the potential is a crucial ingredient in order to bridge the
gap between the physics of disordered and dynamical systems
\cite{tstimp,cit-Mirlin} and to obtain the Lyapunov regime
\cite{cit-Jalabert-Past}. Moreover, taking a finite $\xi$ is not only helpful
for computational or conceptual purposes, but it constitutes an appropriate
approximation for an uncontrolled error in the reversal procedure
$\mathcal{H}_{0}\rightarrow-\mathcal{H}_{0}+\Sigma$ as well as an approximate
description for an external environment. Without entering into a discussion
about what kind of perturbation more appropriately represents an external
environment, it is reasonable to admit that the interaction with the
environment will not be local (or short range), but will extend over certain
typical length.

Another important point to discuss concerning the appropriateness of our
perturbation toward the representation of an external environment, is its time
dependence. Taking a quenched disorder perturbation that only acts in the
second half of the time evolution, represents a very crude approximation to
the dynamics of a more realistic environment \cite{dissipation}. Moreover,
since the disorder is quenched, there is no feed--back of the system on the
environment. In view of our main result, the robustness of the Lyapunov regime
respects to the details of the perturbation, this limitation should not
prevent us of extrapolating our results to realistic cases, and envisioning
their experimental consequences. In any case, a semiclassical approach to time
dependent perturbations shows that the results of Ref.\onlinecite{cit-Jalabert-Past}
remains fairly unchanged \cite{CucLewPas}.

As discussed in the previous chapter, in the leading order of $\hbar$ and for
sufficiently weak perturbations, we can neglect the changes in the classical
dynamics associated with the disorder. We simply modify the contributions to
the semiclassical expansion of the LE associated with a trajectory $s$ (or in
generally to any quantity that can be expressed in terms of the propagators)
by adding the extra phase $\Delta S$ of Eq.~(\ref{eq:DeltaS}). For the
perturbation (\ref{eq:irrevpart}) we can make the change of variables
$\mathbf{q}=\mathbf{v}\bar{t}$ and write%

\begin{equation}
\label{eq:dis_action}\Delta S_{s} \ = \ - \ \frac{1}{v_{0}} \ \int
_{\mathcal{C}_{s}^{\mathrm{c}}} V(\mathbf{q}) \ \mathrm{d}q \ .
\end{equation}

\noindent The integration is now over the unperturbed trajectory
$\mathcal{C}_{s}^{\mathrm{c}}$, and we have assumed that the velocity along
the trajectory remains unchanged respect to its initial value $v_{0}%
=p_{0}/m=L_{s}/t$.

For trajectories of length $L_{s}\gg\xi$, the contributions to $\Delta S$ from
segments separated more than $\xi$ are uncorrelated. The stochastic
accumulation of action along the path can be therefore interpreted as
determined by a random-walk process, resulting in a Gaussian distribution of
$\Delta S_{s}(L_{s})$. This has also been verified numerically in
Ref.~\onlinecite{Heller}. The integration over trajectories represents a for
of average for $\exp[\tfrac{i}{\hbar}\Delta S_{s}].$ The ensemble average over
the propagator (\ref{eq-SemiclasicalPropagator}) (or over independent
trajectories in Eq.~(\ref{eq-MCompleto})) is then obtained from%

\begin{equation}
\langle\exp\left[  \tfrac{\mathrm{i}}{\hbar}\Delta S_{s}\right]  \rangle
=\exp\left[  -\frac{\langle\Delta S_{s}^{2}\rangle}{2\hbar^{2}}\right]  \ ,
\label{eq:gauss_av}%
\end{equation}

\noindent and therefore entirely specified by the variance%

\begin{equation}
\label{eq:dS2}\langle\Delta S_{s}^{2} \rangle= \frac{1}{v_{0}^{2}}
\int_{\mathcal{C}_{s}^{c}} \mathrm{d}q \ \int_{\mathcal{C}_{s}^{c}}
\mathrm{d}q^{\prime}\ \langle V(\mathbf{q}) V(\mathbf{q}^{\prime}) \rangle\ .
\end{equation}

Since the length $L_{s}$ of the trajectory is supposed to be much larger than
$\xi$, the integral over $q-q^{\prime}$ can be taken from $-\infty$ to
$+\infty$, while the integral on $(q+q^{\prime})/2$ gives a factor of $L_{s}$.
We thus have%

\begin{equation}
\label{eq:dS2fr}\langle\Delta S^{2} \rangle= \frac{L_{s}}{v_{0}^{2}}
\ \int\mathrm{d}q \ C(\mathbf{q}) \ ,
\end{equation}

\noindent resulting in

\begin{equation}
\langle\exp\left[  \frac{\mathrm{i}}{\hbar}\Delta S_{s}\right]  \rangle
=\exp\left[  -\frac{L_{s}}{2\tilde{\ell}}\right]  =\exp\left[  -\frac{v_{0}%
t}{2\tilde{\ell}}\right]  \ . \label{eq:gauss_av2}%
\end{equation}

\noindent Where, in analogy with disordered systems \cite{JMP,tstimp}, we have
defined the typical length over which the quantum phase is modified by the
perturbation as%

\begin{equation}
\frac{1}{\tilde{\ell}}=\frac{1}{\hbar^{2}v_{0}^{2}}\ \int\mathrm{d}%
q\ C(\mathbf{q})=\frac{u^{2}n_{i}}{v_{0}^{2}\hbar^{2}(4\pi\xi^{2})^{(d-1)/2}%
}\ . \label{eq:mfp_gauss}%
\end{equation}

\noindent The \textquotedblleft elastic mean free path" $\tilde{\ell}$ and the
mean free time $\tilde{\tau}=\tilde{\ell}/v_{0}^{{}}$ associated with the
perturbation \cite{mistake} will constitute a measure of the strength of the coupling.

Taking impurity averages is technically convenient, but not crucial. Results
like that of Eq.~(\ref{eq:gauss_av2}) would also arrive from considering a
single impurity configuration and a large number of trajectories exploring
different regions of phase space.

\subsection{Loschmidt echo in a classically chaotic system}

\label{subsec:leiaccs}

Once we have settled the hypothesis with respect to the perturbation, we can
go back to Eqs.~(\ref{eq-MCompleto}) and (\ref{eq:decomposition}) calculate
the two contributions to the Loschmidt echo.

In the non-diagonal term the impurity average can be done independently for
$s$ and ${s}^{\prime}$, since the two trajectories explore different regions
of phase space. Therefore, upon impurity average the non-diagonal term becomes%

\begin{equation}
M^{\mathrm{nd}}(t)=\left\vert \left\langle m(t)\right\rangle \right\vert
^{2}=\left(  \frac{\sigma^{2}}{\pi\hbar^{2}}\right)  ^{d}\ \ \left\vert
\ \int\mathrm{d}\mathbf{r}\sum_{s}\ C_{s}\ \exp{\left[  -\frac{\sigma^{2}%
}{\hbar^{2}}\left(  {\overline{\mathbf{p}}}_{s}-\mathbf{p}_{0}\right)
^{2}\right]  }\ \langle\exp\left[  \frac{\mathrm{i}}{\hbar}\Delta
S_{s}\right]  \rangle\right\vert ^{2}\ .
\end{equation}

\noindent We have kept the same notation for the averaged and individual LE,
in order to simplify the notation, and because it will be demonstrated that
this distinction is not crucial. According to Eq.~(\ref{eq:gauss_av2}) we have
\cite{cit-Jalabert-Past}%

\begin{equation}
M^{\mathrm{nd}}(t)=\left(  \frac{\sigma^{2}}{\pi\hbar^{2}}\right)
^{d}\ \ \exp\left[  -\frac{v_{0}t}{\tilde{\ell}}\right]  \ \left\vert
\ \int\mathrm{d}\mathbf{r}\sum_{s}\ C_{s}\ \exp{\left[  -\frac{\sigma^{2}%
}{\hbar^{2}}\left(  {\overline{\mathbf{p}}}_{s}-\mathbf{p}_{0}\right)
^{2}\right]  }\right\vert ^{2}=\exp\left[  -\frac{v_{0}t}{\tilde{\ell}%
}\right]  \ . \label{eq-MNDiagonal}%
\end{equation}

\noindent This term depends on the perturbation, through $\tilde{\ell}$, and
can be interpreted as a Fermi golden rule result \cite{cit-Jacquod01}.

In the diagonal term the trajectories $s$ and $s^{\prime}$ of
Eq.~(\ref{eq-MCompleto}) remain close to each other. The existence of such
types of trajectories is based on the structural stability of the manifold
\cite{cit-CerrutiTomsovic,Heller} (opposed to the exponential sensitivity of
individual trajectories). The actions $\Delta S_{s}$ and $\Delta S_{s^{\prime
}}$ accumulated by effect of the perturbation cannot be taken as uncorrelated,
like in the previous case. A special treatment should be applied to the terms
arising from $s\simeq s^{\prime}$. The small difference between $s$ and
$s^{\prime}$ is only considered through the difference of actions, and therefore%

\begin{equation}
M^{\mathrm{d}}(t)=\left(  \frac{\sigma^{2}}{\pi\hbar^{2}}\right)  ^{d}%
\int\mathrm{d}\mathbf{r}\int\mathrm{d}\mathbf{r}^{\prime}\sum_{s}C_{s}^{2}%
\exp\left[  -\frac{2\sigma^{2}}{\hbar^{2}}\left(  {\overline{\mathbf{p}}}%
_{s}-\mathbf{p}_{0}\right)  ^{2}\right]  \left\langle \exp\left[
\frac{\mathrm{i}}{\hbar}\left(  \Delta S_{s}-\Delta S_{s^{\prime}}\right)
\right]  \right\rangle . \label{eq-MDiagonal}%
\end{equation}

Since $s$ and $s^{\prime}$ are nearby trajectories, we can write%

\begin{equation}
\Delta S_{s}-\Delta S_{s^{\prime}}=\int_{0}^{t}\mathrm{d}~\bar{t}%
\ \ \nabla\tilde{V}(\mathbf{q}_{s}(\bar{t}))\cdot\left(  \mathbf{q}_{s}%
(\bar{t})-\mathbf{q}_{s^{\prime}}(\bar{t})\right)  \ . \label{eq:phachange}%
\end{equation}

\noindent The difference between the intermediate points of both trajectories
can be expressed using the matrix $B$ of Eq.~(\ref{eq:matB}):%

\begin{equation}
\mathbf{q}_{s}(\bar{t})-\mathbf{q}_{s^{\prime}}(\bar{t}) = B^{-1}(\bar{t})
\left(  {\overline{\mathbf{p}}}_{s}-{\overline{\mathbf{p}}}_{s^{\prime}%
}\right)  = B^{-1}(\bar{t})B(t)\left(  \mathbf{r}-\mathbf{r}^{\prime}\right)
\ . \label{eq:operatorB}%
\end{equation}

In the chaotic case the behavior of $B^{-1}(\bar{t})$ is dominated by the
largest eigenvalue $e^{\lambda\bar{t}}$. Therefore we make the simplification
$B^{-1}(\bar{t})B(t)=\exp{\left[  \lambda(\bar{t}-t)\right]  }I$, where $I$ is
the unit matrix and $\lambda$ the mean Lyapunov exponent. Here, we use our
hypothesis of strong chaos which excludes marginally stable
regions\cite{weak-chaos} with anomalous time behavior. Assuming a Gaussian
distribution for the random variable $\Delta S_{s}-\Delta S_{s^{\prime}}$
\cite{Heller}, in analogy with Eq.~(\ref{eq:gauss_av}), we have%

\begin{equation}
\left\langle \exp\left[  \frac{\mathrm{i}}{\hbar}\left(  \Delta S_{s}-\Delta
S_{{s}^{\prime}}\right)  \right]  \right\rangle =\exp\left[  -\frac{1}%
{2\hbar^{2}}\ \int_{0}^{t}\mathrm{d}\bar{t}\int_{0}^{t}\mathrm{d}\bar
{t}^{\prime}\exp\left[  \lambda(\bar{t}+\bar{t}^{\prime}-2t)\right]
\ C_{\nabla\tilde{V}}(|q_{s}(\bar{t})-q_{s}(\bar{t}^{\prime})|)\ \left(
\mathbf{r}-\mathbf{r}^{\prime}\right)  ^{2}\right]  \ . \label{eq:difdels}%
\end{equation}

\noindent Unlike the non-diagonal case, that was obtained through the
correlation potential (Eq.~(\ref{eq:corr})), we are now led to consider the
``force correlator''%

\begin{equation}
C_{\nabla\tilde{V}}(|\mathbf{q}-\mathbf{q}^{\prime}|)=\langle\nabla\tilde
{V}(\mathbf{q})\cdot\nabla\tilde{V}(\mathbf{q}^{\prime})\rangle=\frac
{u^{2}n_{i}}{\left(  4\pi\xi^{2}\right)  ^{d/2}}\left(  \frac{d}{2\xi^{2}%
}-\left(  \frac{\mathbf{q}-\mathbf{q}^{\prime}}{2\xi^{2}}\right)  ^{2}\right)
\exp\left[  -\frac{1}{4\xi^{2}}(\mathbf{q}-\mathbf{q}^{\prime})^{2}\right]
\ .
\end{equation}

\noindent Using the fact that $C_{\nabla\tilde{V}}$ is short-ranged (in the
scale of $\xi$), and working in the limit $\lambda t \gg1$, the integrals of
Eq.~(\ref{eq:difdels}) yield%

\begin{equation}
\left\langle \exp\left[  \frac{\mathrm{i}}{\hbar}\left(  \Delta S_{s}-\Delta
S_{s^{\prime}}\right)  \right]  \right\rangle =\exp\left[  -\frac{A}%
{2\hbar^{2}}\left\vert \mathbf{r}-\mathbf{r}^{\prime}\right\vert ^{2}\right]
\ . \label{eq:deltmdelsp}%
\end{equation}

\noindent with%

\begin{equation}
A=\frac{(d-1)u^{2}n_{i}}{4\lambda v_{0}\xi^{2}(4\pi\xi^{2})^{(d-1)/2}} \ .
\label{eq:Aquencheddis}%
\end{equation}

\noindent Therefore, we have%

\[
M^{\mathrm{d}}(t)=\left(  \frac{\sigma^{2}}{\pi\hbar^{2}}\right)  ^{d}%
\int\mathrm{d}\mathbf{r}\int\mathrm{d}\mathbf{r}^{\prime}\ \sum_{s}\ C_{s}%
^{2}\ \exp\left[  -\frac{2\sigma^{2}}{\hbar^{2}}\left(  {\overline{\mathbf{p}%
}}_{s}-\mathbf{p}_{0}\right)  ^{2}\right]  \ \exp{\left[  -\frac{A}{2\hbar
^{2}}\left(  \mathbf{r}-\mathbf{r}^{\prime}\right)  ^{2}\right]  }\ .
\]

\noindent A Gaussian integration over $(\mathbf{r}-\mathbf{r}^{\prime})$
results in%

\[
M^{\mathrm{d}}(t)=\left(  \frac{\sigma^{2}}{\pi\hbar^{2}}\right)  ^{d}%
\ \int\mathrm{d}\mathbf{r}\ \sum_{s}\ C_{s}^{2}\ \left(  \frac{2\pi\hbar^{2}%
}{A}\right)  ^{d/2}\ \exp\left[  -\frac{2\sigma^{2}}{\hbar^{2}}\left(
{\overline{\mathbf{p}}}_{s}-\mathbf{p}_{0}\right)  ^{2}\right]  \ .
\]

\noindent The factor $C_{s}^{2}$ reduces to $C_{s}$ when we make the change of
variables from $\mathbf{r}$ to ${\overline{\mathbf{p}}}$. In the long-time
limit $C_{s}^{-1}\propto e^{\lambda t}$, while for short times $C_{s}%
^{-1}=(t/m)^{d}$. Using a form that interpolates between these two limits we have%

\begin{equation}
M^{\mathrm{d}}(t)=\left(  \frac{\sigma^{2}}{\pi\hbar^{2}}\right)  ^{d}%
\int\mathrm{d}{\overline{\mathbf{p}}}\ \left(  \frac{2\pi\hbar^{2}}{A}\right)
^{d/2}\left(  \frac{m}{t}\right)  ^{d}\ \exp\left[  -\lambda t\right]
\ \exp\left[  -\frac{2\sigma^{2}}{\hbar^{2}}\left(  {\overline{\mathbf{p}}%
}-\mathbf{p}_{0}\right)  ^{2}\right]  =\overline{A}\exp\left[  -\lambda
t\right]  \ , \label{eq:oversqdi3}%
\end{equation}

\noindent with $\overline{{A}}=[\sigma m/(A^{1/2}t)]^{d}$. Since the integral
over ${\overline{\mathbf{p}}}$ is concentrated around $\mathbf{p}_{0}$, the
exponent $\lambda$ is taken as the phase-space average value on the
corresponding energy shell. The coupling $\Sigma$ appears only in the
prefactor (through $\overline{{A}}$) and therefore its detailed description is
not crucial in discussing the time dependence of $M^{\mathrm{d}}$.

The limits of small $t$ and weak $\Sigma$ yield an infinite $\overline{{A}}$,
and thus a divergence in Eq.~(\ref{eq:oversqdi3}). However, our calculations
are only valid in certain intervals of $t$ and strength of the perturbation.
The times considered should verify $v_{0}t/\tilde{\ell}\geq1$. \ Long times,
resulting in the failure of our diagonal approximations
(Eqs.~(\ref{eq-MCompleto}) and (\ref{eq-MDiagonal})) or our assumption that
the trajectories are unaffected by the perturbation, are excluded from our
analysis. Similarly, the small values of $\Sigma$ are not properly treated in
the semiclassical calculation of the diagonal term $M^{\mathrm{d}}(t)$, while
for strong $\Sigma$ the perturbative treatment of the actions is expected to
break down and the trajectories are affected by the quenched disorder. This
last condition translates into a \textquotedblleft transport
mean-free-path\textquotedblright\cite{JMP,tstimp} $\tilde{\ell}_{\mathrm{tr}%
}=4(k\xi)^{2}\tilde{\ell}$ much larger than the typical dimension $R$ of our
system. In the limit $k\xi\gg1$ that we are working with, we are able to
verify the condition $\tilde{\ell}_{\mathrm{tr}}\gg R\gg\tilde{\ell}$.

Within the above limits, our semiclassical approach made it possible to
estimate the two contributions of Eq.~(\ref{eq:decomposition}) to $M(t)$. The
non-diagonal component $M^{\mathrm{nd}}(t)$ will dominate in the limit of
small $t$ or $\Sigma$. In particular, such a contribution ensures that
$M_{\Sigma=0}(t)=1$ (see Eq.~(\ref{eq:mwse0})), and that $M_{\Sigma
}(t\!=\!0)=1$. The diagonal term will dominate over the non-diagonal one for
perturbations strong enough to verify%

\begin{equation}
\tilde{\ell}<\frac{v_{0}}{\lambda}\ . \label{eq:condition}%
\end{equation}

\noindent This crossover condition is extremely important, and will be
discussed in detail in the sequel.

It is worth to notice that the width $\sigma$ of the initial wave-packet 
is a prefactor of the
diagonal contribution. The non-diagonal term, on the other hand, is
independent on the initial wave-packet. Therefore, as explained in
Ref.~\onlinecite{cit-JacquodSP}, changing our initial state
(\ref{eq-PaqueteInicial}) into a coherent superposition of $N$ wave-packets
would reduce $M^{\mathrm{d}}$ by a factor of $N$ without changing
$M^{\mathrm{nd}}$. The localized character of the initial state is then a key
ingredient in order to obtain the universal behavior.

\section{Loschmidt echo in the two-dimensional Lorentz gas}

\label{sec:lorentz}

\subsection{Classical dynamics of $\mathcal{H}_{0}$}

\label{sec:classproplorentz}

We consider in this section the case where the system Hamiltonian
$\mathcal{H}_{0}$ represents a two dimensional Lorentz gas, i.e. a particle
that moves freely (with speed $v$) between elastic collisions (with specular
reflections) with an irregular array of hard disk scatterers (impurities) of
radius $R$. Such a billiard system is a paradigm of classical dynamics, and
has been proven to exhibit mixing and ergodic behavior, while its dynamics for
long distances is diffusive \cite{cit-Arnold,Dorfmanbook,cit-AleinerLarkin}.
The existence of rigorous results for the Lorentz gas has made of it a
preferred playground to study the emergence of irreversible behavior out of
the reversible laws of classical dynamics \cite{Dorfmanbook}. Moreover,
anti-dot lattices defined in a two dimensional electron gas
\cite{cit-Weiss,cit-Richter,cit-Portal} constitute an experimentally
realizable quantum system where classical features have been identified and
measured. We will use the terms anti-dot and disk indistinctly.

In our numerical simulations we are limited to finite systems, therefore we
will work in a square billiard of area $L^{2}$ (with $N$ scatterers), and
impose periodic boundary conditions. The concentration of disks is
\begin{equation}
c=N\pi R^{2}/L^{2}.
\end{equation}
We require that each scatterer has an exclusion region $R_{\mathrm{e}}$ from
its border, such that the distance between the centers of any pair of disks is
larger than a value $2R_{\mathrm{e}}>2R.$ Such a requirement is important to
avoid the trapping of the classical particle and the wave-function
localization in the quantum case. The anti-dots density is set to be roughly
uniform, and the concentration is chosen to be the largest one compatible with
the value of $R_{\mathrm{e}}$, obtained numerically as $c=0.7\pi
R^{2}/4R_{\mathrm{e}}^{2}.$

The Lorentz gas has been thoroughly studied\cite{Dorfmanbook}, and we will not
discuss here its classical dynamics in detail. We will simply recall some of
its properties that will be used in the sequel, and present the numerical
simulations that allow us to extract some important physical parameters.

\begin{figure}[tb]
\centering \includegraphics*[width=8.3cm]{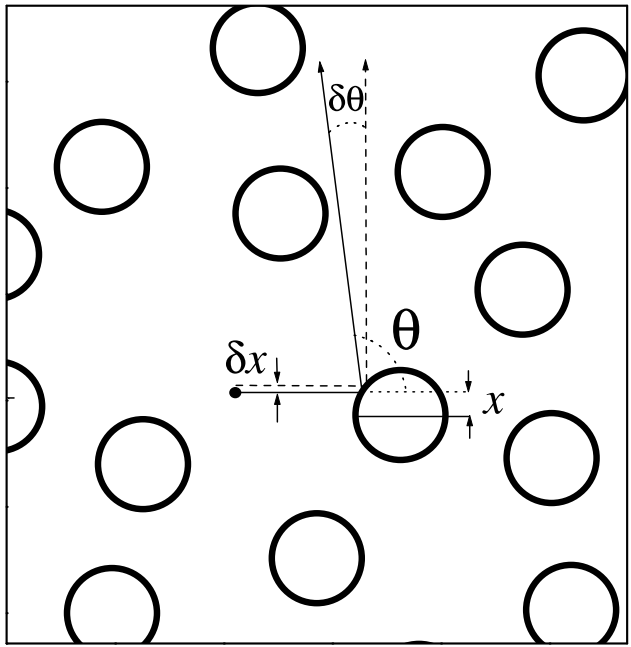}\caption{Schematics of a
Lorentz gas showing the dispersion of two trajectories initially close to each
other (with a difference $\delta x$ in the impact parameter $x$). The angle
$\delta\theta$ between the two trajectories increases after each collision as
described in the text.}%
\label{fig-ColisionAnalitica}%
\end{figure}

The chaotic character of the dynamics is a consequence of the de-focusing
nature of the collisions. As illustrated in Fig.~\ref{fig-ColisionAnalitica},
a particle with impact parameter $x$ will be reflected with an angle
\begin{equation}
\theta=\pi-2\arctan\left[  \frac{x}{\sqrt{R^{2}-x^{2}}}\right]  .
\end{equation}
If we consider a second particle with impact parameter $x+\delta x,$ its
outgoing angle will be $\theta+\delta\theta,$ with
\begin{equation}
\delta\theta=\frac{2}{\sqrt{R^{2}-x^{2}}}\ \delta x.
\end{equation}
The separation between the two particles that have traveled a distance $s$
after the collision will grow as
\begin{equation}
\delta d\simeq\delta x+\delta\theta s\simeq\delta x\left(  1+\frac{2s}%
{\sqrt{R^{2}-x^{2}}}\right)  . \label{eq-Separation}%
\end{equation}
The next collision will further amplify the separation, due to the new impact
parameters and the different incidence angles.

Within the above restrictions, the exclusion distance $R_{\mathrm{e}}$
completely determines the dynamical properties of the Lorentz gas. Among them,
we are interested in the Lyapunov exponent (measuring the rate of separation
of two nearby trajectories), the elastic mean free path $\ell$ (given by the
typical distance between two collisions), and the transport mean free path
$\ell_{\mathrm{tr}}$ (defined as the distance over which the momentum is
randomized and the dynamics can be taken as effectively diffusive).

\begin{figure}[tb]
\centering \includegraphics*[width=8.3cm]{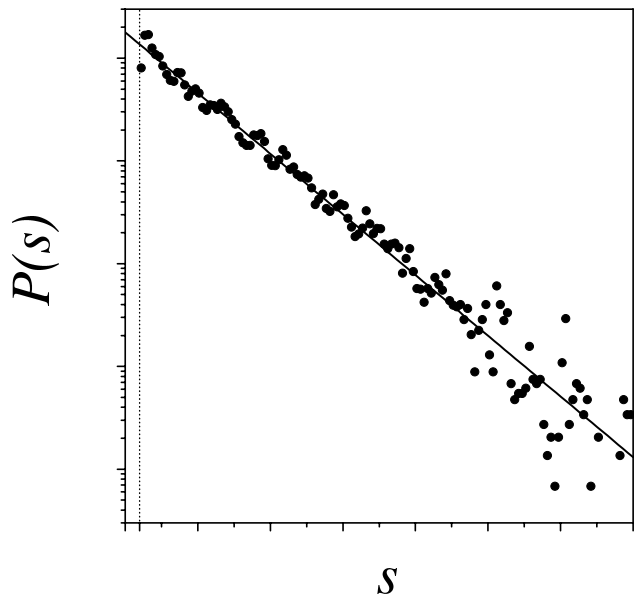}
\caption{Histogram of the distances between collisions with the disks, used in
order to obtain numerically the mean free path $\ell$ for the Lorentz gas. The
solid line represents Eq.~(\ref{eq-DistributionL}) and the dash-dotted
vertical line the cut-off distance $2(R_{\mathrm{e}}-R)$.}%
\label{fig-ShiftedPoisson}%
\end{figure}

A shifted Poisson distribution%

\begin{equation}
P(s)=\left\{
\begin{array}
[c]{cl}%
{\displaystyle{\frac{\exp\left[  -s/(\ell-2(R_{\mathrm{e}}-R))\right]  }%
{(\ell-2(R_{\mathrm{e}}-R))\exp\left[  -2(R_{\mathrm{e}}-R)/(\ell
-2(R_{\mathrm{e}}-R))\right]  }}} & \hspace{1cm}%
\mbox{if $s>2(R_{\rm e}-R)$}\ ,\\
0 & \hspace{1cm}\mbox{if $s<2(R_{\rm e}-R)$}\ ,
\end{array}
\right.  \label{eq-DistributionL}%
\end{equation}

\noindent is a reasonable guess for the distribution of lengths between
successive collisions, which yields $\left\langle s\right\rangle =\ell
=v/\tau_{\mathrm{e}}$, and is consistent with numerical simulations in the
range of anti-dot concentration that we are interested in (see
Fig.~\ref{fig-ShiftedPoisson}). Since velocity both\ $v_{0}$ and momentum
$p_{0}$\ are conserved within this model we will drop their subindex.

The elastic mean free path that we obtain from Fig.~\ref{fig-ShiftedPoisson}
compares favorably with a simple estimation of the mean free distance in a
strip of length $L$ and width $2R$ with $2cL/\pi R$ disks,%

\begin{equation}
\ell\simeq\frac{\pi R}{2c} \ - \ \frac{\pi R}{2} = \frac{R_{\mathrm{e}}^{2}%
}{0.35R} \ - \ \frac{\pi R}{2}.
\end{equation}

The diffusive character of the Lorentz gas can be put in evidence from the
time evolution of the root mean square displacement over a collection of
initial conditions. We numerically obtain $\left\langle r^{2}(t)\right\rangle
=2dDt$ (with $d=2$). $\tau_{\mathrm{tr}}=\ell_{\mathrm{tr}%
}/v$ is the mean time required to randomize the direction and $D=v\ell
_{\mathrm{tr}}/2d$ is the diffusion coefficient. The difference between $\ell$
and $\ell_{\mathrm{tr}}$ arises from the angular dependence of the scattering
cross section. Taking this factor into account we obtain a ratio
$\ell_{\mathrm{tr}}/\ell$ which is in good agreement with the one obtained
from the independently determined $\ell$ and $\ell_{\mathrm{tr}}.$

There are known various estimations of the Lyapunov exponent of the Lorentz
gas in different regimes. Considering the three-disk problem, Gaspard \emph{et
al}. \cite{cit-Gaspard} obtained%

\begin{equation}
\lambda=\frac{v}{2R_{\mathrm{e}}-2R}\ \ln\left[  \frac{2R_{\mathrm{e}%
}-R+\left(  4R_{\mathrm{e}}^{2}-4R_{\mathrm{e}}R\right)  ^{1/2}}{R}\right]
\ .
\end{equation}
Considering a periodic Lorentz gas (repeated Sinai billiard) Laughlin proposed
the form \cite{cit-Laughlin}
\begin{equation}
\lambda=\frac{v}{\ell}\ \ln\left[  1+\frac{\beta\ell}{R}\right]  \ ,
\label{eq-Laughlin}%
\end{equation}

\noindent where $\beta$ is a geometrical factor of order $1.$ In the diluted
limit ($c\ll1$), van Beijeren and Dorfman \cite{cit-BeijerenDorfman} showed that%

\begin{equation}
\lambda=2\ \frac{N}{L^{2}}\ Rv\left(  1-\ln2-0.577-\ln\left[  \frac{NR^{2}%
}{L^{2}}\right]  \right)  .
\end{equation}

\begin{figure}[tb]
\centering \includegraphics*[width=8.3cm]{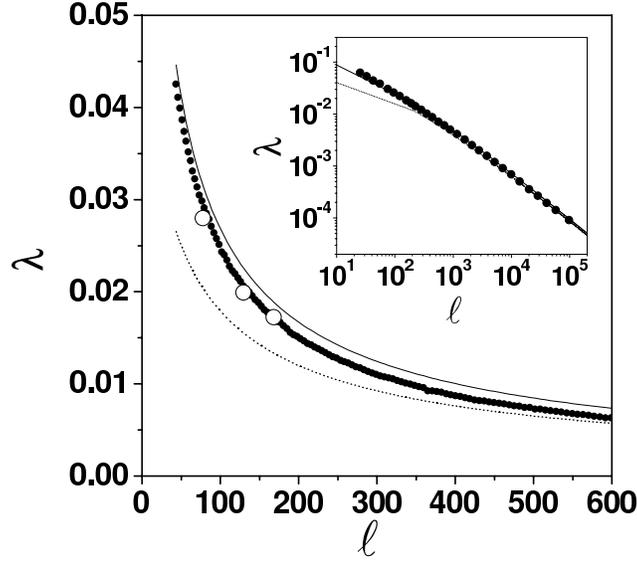}
\caption{Lyapunov exponent $\lambda$ of the Lorentz gas as a function of the mean free
path $\ell$. The black dots represent our numerical values and the solid line
the analytical estimate of Eq.~(\ref{eq-LyapunovAnalitico}). The dashed line
indicates Laughlin's approximation (Eq.~(\ref{eq-Laughlin}) and the open
dots are the quantum values obtained from the decay of the LE
(Fig.~\ref{fig-TauPhivsAlfa} in the sequel). Inset: the same plot in log-log
scale highlighting the agreement between the different approximations in the
region of very small concentrations (large $\ell$).}%
\label{fig-LyapunovAnalitico}%
\end{figure}

Numerically, the procedure of Benettin et al. \cite{cit-Lyapunov} is usually
followed in order to obtain Lyapunov exponents. Two nearby trajectories are
followed, and their separation is scaled down to the initial value $\delta
x_{0}$ after a characteristic time $t$ (that we take it to be larger than the
collision time). We can then limit the numerical errors, and avoid entering
the diffusive regime where two initially close trajectories follow completely
independent paths. The Lyapunov exponent results from the average over the
expanding rates in the different intervals,
\begin{equation}
\lambda=\lim_{n\rightarrow\infty}\frac{v}{n}\ \sum_{j=1}^{n}\frac{1}{s_{j}%
}\ \ln\left[  \frac{\delta x_{j}}{\delta x_{0}}\right]  \ ,
\end{equation}
where $s_{j}$ is the length of the $j$-th interval, and $\delta x_{j}$ the
separation just before the normalization. Technically, we should work with
distances in phase-space, rather than in configuration space, but the local
instability makes this precision unnecessary.

Benettin's algorithm can also be used for a semi-analytical calculation of the
Lyapunov exponent. Taking the length distribution of
Eq.~(\ref{eq-DistributionL}) to obtain the average separation after a
collision from Eq.~(\ref{eq-Separation}), and identifying the average over
pieces of the trajectory with a geometrical average over impact parameters, we
can write
\begin{equation}
\lambda=\frac{v}{R \ell} \ \int_{0}^{R} dx \ \ln{\left[  1+\frac{2 \ell}%
{\sqrt{R^{2}-x^{2}}}\right]  } \ .
\end{equation}

Performing the integration yields
\begin{equation}
\frac{\lambda}{v} =\frac{1}{\ell} \ \ln{\left[  \frac{\ell}{R}\right]  }+
\frac{\pi}{R}+\sqrt{\frac{4}{R^{2}} - \frac{1}{\ell^{2}}} \ \left(
\arcsin{\left[  \frac{R}{2 \ell}\right]  } - \frac{\pi}{2} \right)  \ .
\label{eq-LyapunovAnalitico}%
\end{equation}
As shown in Fig.~\ref{fig-LyapunovAnalitico}, the above expression reproduces
remarkably well the numerical calculations of the Lyapunov exponent. It agrees
also with the result of van Beijeren and Dorfman in the dilute limit, and
gives good agreement with Laughlin's estimation.

\subsection{Perturbation Hamiltonian}

The quantum and classical fidelity measure the sensitivity of a given system
to a perturbation of its Hamiltonian. In Ref.~\onlinecite{cit-Jalabert-Past} a
quenched disorder environment was taken as the perturbation and the relaxation
rate was found to depend only on the system Hamiltonian; the details of the
perturbation are not important beyond some critical strength. Subsequent works
have tried this perturbation \cite{cit-SmoothBilliard} and others
\cite{cit-PhysicaA}$^{-}$\cite{cit-Mirlin} confirming the universality of the
result of Ref. \onlinecite{cit-Jalabert-Past}. It is also useful to verify
such an universality by considering completely different perturbations on a
given system. The Lorentz gas is an ideal case, since it can be perturbed by a
quenched disorder or by the distortion of the mass tensor, introduced in
Ref.~\onlinecite{cit-Lorentzgas} and briefly discussed in the sequel.

The isotropic mass tensor of $\mathcal{H}_{0},$ of diagonal components
$m_{0},$ can be distorted by introducing an anisotropy such that $m_{xx}%
=m_{0}(1+\alpha)$ and $m_{yy}=m_{0}/(1+\alpha).$ This perturbation is inspired
by the effect of a slight rotation of the sample in the problem of dipolar
spin dynamics \cite{cit-QZeno}, which modifies the mass of the spin wave
excitations. The kinetic part of the Hamiltonian is now affected by the
perturbation, that writes as
\begin{equation}
\Sigma(\alpha)=\alpha\ \frac{p_{y}^{2}}{2m_{0}}-\frac{\alpha} {1+\alpha}
\ \frac{p_{x}^{2}}{2m_{0}}. \label{eq-Perturbation}%
\end{equation}

\noindent In our analytical work we will stay within the leading order
perturbation in $\alpha$. That is,
\begin{equation}
\Sigma(\alpha)=\frac{\alpha}{2m_{0}} \ \left(  p_{y}^{2}-p_{x}^{2}\right)  .
\label{eq-PerturbationAprox}%
\end{equation}

Making the particle \textquotedblleft heavier\textquotedblright\ in the $x$
direction (i.e. we consider a positive $\alpha)$ modifies the equations of
motion without changing the potential part of the Hamiltonian. It is important
to notice that, unlike the case of quenched disorder, the perturbation
(\ref{eq-Perturbation}) is non-random, and will not be able to provide any
averaging procedure by itself, but through the underlying chaotic dynamics.

Numerical simulations of the evolution of two trajectories with the same
initial conditions, the first one governed by $\mathcal{H}_{0}$ and the second
one by $\mathcal{H}_{0}+\Sigma,$ show that the distance in phase space grows
exponentially with the same Lyapunov exponent that amplifies initial
distances. The classical dynamics is then equally sensitive to changes in the
Hamiltonian as to changes in the initial conditions \cite{cit-Shack-Caves}.

For a hard wall model, like the one we are considering, the perturbation
(\ref{eq-Perturbation}) is equivalent to having non-specular reflections. One
can resort to the minimum-action principle (see appendix \ref{ape:reflexion})
to obtain a generalized reflection law:%

\begin{subequations}
\label{eq-ReflectionLaw}%
\begin{align}
v_{x}^{\prime}  &  =\frac{v_{x}(m_{x}n_{y}^{2} -m_{y}n_{x}^{2})-2v_{y}%
m_{y}n_{x}n_{y}} {m_{x}n_{y}^{2}+m_{y}n_{x}^{2}} \ , \label{eq-ReflectionLaw0}%
\\
\displaystyle v_{y}^{\prime}  &  =\frac{v_{y}(m_{y}n_{x}^{2}- m_{x}n_{y}%
^{2})-2v_{x}m_{x}n_{x}n_{y}}{m_{x}n_{y}^{2}+m_{y}n_{x}^{2}} \ .
\label{eq-ReflectionLaw1}%
\end{align}

\noindent where $v_{x}^{\prime}$ and $v_{y}^{\prime}$ are the two components
of the velocity after a collision against a surface defined by its normal
unitary vector $(n_{x},n_{y}).$ Eqs.~(\ref{eq-ReflectionLaw}) allow to show
that the distortion of the mass tensor is equivalent to an area conserving
deformation of the boundaries as $x\rightarrow x(1+\xi)$, $y\rightarrow
y/(1+\xi),$ as used in other works on the LE\cite{cit-Bunimovich}, where
$\xi=\sqrt{1+\alpha}-1$ is the stretching parameter. An illustrative example
of the equivalence of the perturbations is shown for a stretched stadium
billiard in Fig.~\ref{fig-RelacionPerturbaciones}. In panel $a$, a typical
trajectory is shown; while in panels $b$ and $c$ we see the trajectories
resulting of a perturbation of the mass tensor or the stretch of the
boundaries respectively.

\begin{figure}[tb]
\centering \includegraphics*[height=8.3cm,angle=-90]{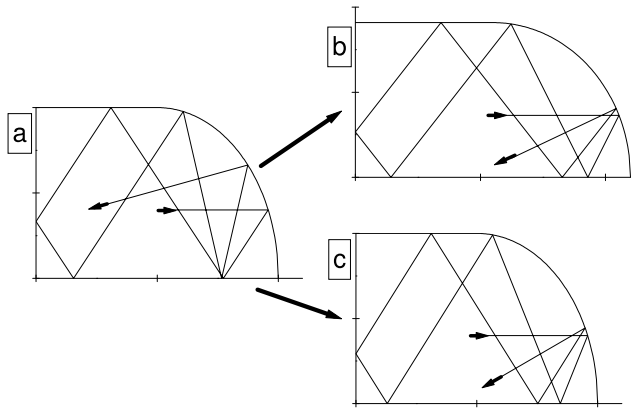}
\caption{A single typical trajectory on the (hard wall) Bunimovich stadium,
the arrows indicate the initial and final positions of the particle. (a)
Normal stadium (b) Stretched stadium. (c) Normal stadium and particle with a
perturbed mass, Eqs. (\ref{eq-Perturbation}). The strengths of the stretching
and the mass distortion are such that the initial condition renders the same
trajectory in both cases, illustrating the corresponding map (see text).}
\label{fig-RelacionPerturbaciones}
\end{figure}

\subsection{Semiclassical Loschmidt echo}

We calculate in this chapter the Loschmidt echo for the system whose classical
counterpart was previously discussed; $\mathcal{H}_{0}$ describes a Lorentz
gas and $\Sigma$ is given by Eq.~(\ref{eq-Perturbation}). We adapt to the
present perturbation the semiclassical approach of Sect.~\ref{sec:semi} and
Ref. \onlinecite{cit-Jalabert-Past}. As before, we take as initial state a
Gaussian wave-packet of width $\sigma$ (Eq.~(\ref{eq-PaqueteInicial})).

The semiclassical approach to the LE under a weak perturbation $\Sigma$ is
given by Eq.~(\ref{eq-MCompleto}), with the extra phase%

\end{subequations}
\begin{equation}
\Delta S_{s}=\int_{0}^{t}\mathrm{d}\overline{t}\ \Sigma_{s}(\mathbf{q(}%
\overline{t}),\mathbf{\dot{q}(}\overline{t})).
\end{equation}

\noindent The sign difference with Eq.~(\ref{eq:DeltaS}) is because the
perturbation is now in the kinetic part of the Hamiltonian. On the other hand,
as explained before, this sign turns out to be irrelevant.

With the perturbation of Eq.~(\ref{eq-PerturbationAprox}) we have to integrate
a piecewise constant function (in between collisions with the scatterers),
obtaining
\begin{equation}
\Delta S_{s}=\frac{\alpha m_{0}}{2} \ \sum_{i=1}^{N_{s}}\tau_{i}\left(
2v_{y_{i}}^{2}-v^{2}\right)  . \label{eq-DeltaAction}%
\end{equation}
We have used $v_{x}^{2}+v_{y}^{2}=v^{2},$ and have defined $\tau_{i}$ as the
free flight time finishing with the $i$-th collision, $v_{y_{i}}$ is the $y$
component of the velocity in such an interval, and $N_{s}$ as the number of
collisions that suffers the trajectory $s$ during the time $t$.

As we saw in chapter \ref{sec:classproplorentz}, the free flight times
$\tau_{i}$ (or the inter-collision length $v\tau_{i})$ have a shifted Poisson
distribution (Eq.~(\ref{eq-DistributionL}) and Fig.~\ref{fig-ShiftedPoisson}).
This observation will turn out to be important in the analytical calculations
that follow since the sum of Eq.~(\ref{eq-DeltaAction}) for a long trajectory
can be taken as composed of uncorrelated random variables following the above
mentioned distribution. It is important to remark that, unlike the case of
Sec.~\ref{sec:semi}, the randomness is not associated with the perturbation
(which is fixed), but with the diffusive dynamics generated by $\mathcal{H}%
_{0}$.

\subsection{Non-diagonal contribution}

As in the case of Sec.~\ref{sec:semi}, the non-diagonal contribution is given
by the second moment
\begin{equation}
\left\langle \Delta S_{s}^{2}\right\rangle =\frac{\alpha^{2}m_{0}^{2}}{4}
\ \left\langle \sum_{i,j=1}^{N_{s}}\tau_{i}\tau_{j}\left(  2v_{y_{i}}^{2}-
v^{2}\right)  \left(  2v_{y_{j}}^{2}-v^{2}\right)  \right\rangle \ .
\end{equation}
Separating in diagonal ($i=j$) and non-diagonal ($i\neq j$) contributions (in
pieces of trajectory) we have
\begin{equation}
\left\langle \Delta S_{s}^{2}\right\rangle =\frac{\alpha^{2}m_{0}^{2}N_{s}}%
{4}\left[  \left\langle \tau_{i}^{2}\right\rangle \left(  4\left\langle
v_{y_{i}}^{4}\right\rangle -4 v^{2}\left\langle v_{y_{i}}^{2}\right\rangle
+v^{4}\right)  +\left(  N_{s}-1\right)  \left\langle \tau_{i}%
\right\rangle ^{2}\left(  4\left\langle v_{y_{i}}^{2}\right\rangle ^{2}%
-4v^{2}\left\langle v_{y_{i}}^{2}\right\rangle +v^{4}\right)  \right]  .
\end{equation}
We have assumed that different pieces of the trajectory ($i\neq j$) are
uncorrelated, and that within a given piece $i,$ $\tau_{i}$ and $v_{y_{i}}$
are also uncorrelated. According to the distribution of time-of-flights
(\ref{eq-DistributionL}) we have%

\begin{subequations}
\label{allTAU}%
\begin{align}
\left\langle \tau\right\rangle  &  =\tau_{\mathrm{e}} \ ,\label{eq:TAU0}\\
\displaystyle \left\langle \tau^{2}\right\rangle  &  =2\tau_{\mathrm{e}}^{2}
\ . \label{eq:TAU1}%
\end{align}

Assuming that the velocity distribution is isotropic ($P(\theta)=1/2\pi$,
where $\theta$ is the angle of the velocity with respect to a fixed axis) is
in good agreement with our numerical simulations, and results in%

\end{subequations}
\begin{subequations}
\label{allVELY}%
\begin{align}
\left\langle v_{y}^{2}\right\rangle  &  =v^{2}\left\langle \sin^{2}%
\theta\right\rangle =\frac{v^{2}}{2} \ ,\label{eq:VELY0}\\
\displaystyle \left\langle v_{y}^{4}\right\rangle  &  =v^{4}\left\langle
\sin^{4}\theta\right\rangle =\frac{3v^{4}}{8} \ . \label{eq:VELY1}%
\end{align}

We thus obtain that $4\left\langle v_{y_{i}}^{2}\right\rangle ^{2}%
-4v^{2}\left\langle v_{y_{i}}^{2}\right\rangle +v^{4}=0,$ implying a
cancellation of the cross terms of $\left\langle \Delta S_{s}^{2}\right\rangle
$, consistently with the lack of correlations between different pieces that we
have assumed. We therefore have
\end{subequations}
\begin{equation}
\left\langle \Delta S_{s}^{2}\right\rangle =\frac{\alpha^{2}m_{0}^{2}N_{s}%
\tau_{\mathrm{e}}^{2}v^{4}}{4} \ .
\end{equation}
For a given $t,$ $N_{s}$ is also a random variable, but for $t\gg
\tau_{\mathrm{e}}$ we can approximate it by its mean value $t/\tau
_{\mathrm{e}}$ and write
\begin{equation}
\left\langle \Delta S_{s}^{2}\right\rangle =\frac{\alpha^{2}m_{0}^{2}v^{4}%
\tau_{\mathrm{e}} t}{4} \ .
\end{equation}
We therefore have for the average echo amplitude%

\begin{equation}
\left\langle m(t)\right\rangle \simeq\exp{\left[  -\frac{\alpha^{2}m_{0}%
^{2}v^{4}\tau_{\mathrm{e}}t}{8\hbar^{2}}\right]  }\left(  \frac{\sigma^{2}%
}{\pi\hbar^{2}}\right)  ^{d/2}\int d\mathbf{r}\sum_{s}C_{s}\exp{\left[
-\frac{\sigma^{2}}{\hbar^{2}}\left(  {\overline{\mathbf{p}}}_{s}%
-\mathbf{p}_{0}\right)  ^{2}\right]  }=\exp{\left[  -\frac{vt}{2\tilde{\ell}%
}\right]  }\ , \label{eq-AmplitudAverage}%
\end{equation}
where we have again used $C_{s}$ as a Jacobian of the transformation from
$\mathbf{r}$ to ${\overline{\mathbf{p}}}_{s}$ and we have defined an effective
mean free path of the perturbation by
\begin{equation}
\frac{1}{\tilde{\ell}}=\frac{m_{0}^{2}v^{2}\ell}{4\hbar^{2}}\ \alpha^{2}\ .
\label{eq-FGRexponent}%
\end{equation}

The effective mean free path $\tilde{\ell}=v~\tilde{\tau}$ should be
distinguished from $\ell=v\tau_{\mathrm{e}}$ since the former is associated to
the dynamics of $\Sigma$ and $\mathcal{H}_{0}$, while the latter is only fixed
by $\mathcal{H}_{0}$. Obviously, our results are only applicable in the case
of a weak perturbation verifying $\tilde{\ell}\gg\ell$. From
Eq.~(\ref{eq-AmplitudAverage}) we obtain the non-diagonal component of the LE
as
\begin{equation}
M^{\mathrm{nd}}(t)=\left\vert \left\langle m(t)\right\rangle \right\vert
^{2}=\exp\left[  -\frac{vt}{\tilde{\ell}}\right]  \ .
\end{equation}

In the next chapters we study the conditions under which the correlations not
contained in the FGR approximation dominate the LE, while in
Sec.~\ref{sec:universality} we will test the above results against numerical simulations.

\subsection{Diagonal contribution}

As in Sec.~\ref{sec:semi}, we have to discuss separately the contribution to
the LE (Eq.~(\ref{eq-MCompleto})) originated by pairs of trajectories $s$ and
$s^{\prime}$ that remain close to each other. In that case the terms $\Delta
S_{s}$ and $\Delta S_{s^{\prime}}$ are not uncorrelated. The corresponding
diagonal contribution to the LE is given by Eq.~(\ref{eq-MDiagonal}), and then
we have to calculate the extra actions for $s\simeq s^{\prime}$. As in
Fig.~\ref{fig-ColisionAnalitica}, we represent by $\theta$ ($\theta+\delta$)
the angle of the trajectory $s$ ($s^{\prime}$) with a fixed direction (i.e.
that of the $x$-axis). We can then write the perturbation
(Eq.~(\ref{eq-Perturbation})) for each trajectory as%

\begin{subequations}
\label{allSIG}%
\begin{align}
\Sigma_{s}  &  =\frac{\alpha}{2m_{0}} \ p^{2} \ (2\sin^{2}\theta-1)
\ ,\label{eq:SIG0}\\
\displaystyle \Sigma_{s^{\prime}}  &  =\frac{\alpha}{2m_{0}} \ p^{2}
\ (2\sin^{2}\theta-2\delta\sin2\theta-1)+\mathcal{O}(\delta^{2}) \ .
\label{eq:SIG1}%
\end{align}

Assuming that the time-of-flight $\tau_{i}$ is the same for $s$ and
$s^{\prime}$ we have
\end{subequations}
\begin{equation}
\Delta S_{s}-\Delta S_{s^{\prime}}=\frac{\alpha p^{2}}{m_{0}} \int_{0}^{t}d
\overline{t} \ \delta(\overline{t}) \ \sin{\left[  2\theta(\overline
{t})\right]  } \ .
\end{equation}

The angles $\delta$ alternate in sign, but the exponential divergence between
nearby trajectories allows to approximate the angle difference after $n$
collisions as $\left|  \delta_{n}\right|  =\left|  \delta_{1}\right|
e^{\lambda n\tau_{\mathrm{e}}}$. A detailed analysis of the classical dynamics
shows that the distance between the two trajectories grows with the number of
collisions as $d_{1}=\left|  \delta_{1}\right|  v\tau_{1},$ $d_{2}%
=d_{1}+\left|  \delta_{2}\right|  v\tau_{2},$ and therefore%

\begin{equation}
d_{N_{s}} \simeq v\sum_{j=1}^{N_{s}}\left|  \delta_{j}\right|  \tau_{j} \simeq
v\tau_{\mathrm{e}}\left|  \delta_{1}\right|  \ \sum_{j=1}^{N_{s}%
}e^{(j-1)\lambda\tau_{\mathrm{e}}} = \ell\ \left|  \delta_{1}\right|
\ \frac{e^{N_{s}\lambda\tau_{\mathrm{e}}}-1} {e^{\lambda\tau_{\mathrm{e}}}-1}
\ .
\end{equation}

\noindent By eliminating $\left|  \delta_{1}\right|  $ we can express an
intermediate angle $\delta(\overline{t})$ as a function of the final
separation $\left|  \mathbf{r}-\mathbf{r}^{\prime}\right|  =d_{N_{s}}$,
\begin{equation}
\delta(\overline{t})\simeq\frac{\left|  \mathbf{r}-\mathbf{r}^{\prime}\right|
}{\ell} \ \frac{e^{\lambda\tau_{\mathrm{e}}}-1}{e^{\lambda t}-1}%
e^{\lambda\overline{t}} \ ,
\end{equation}
where again we have used that $t=N_{s}\tau_{\mathrm{e}}$ is valid on average.
Assuming that the action difference is a Gaussian random variable, in the
evaluation of Eq.~(\ref{eq-MDiagonal}) we only need its second moment%

\begin{equation}
\left\langle \left(  \Delta S_{s}-\Delta S_{s^{\prime}}\right)  ^{2}%
\right\rangle \simeq\alpha^{2}m_{0}^{2}v^{4}\ \frac{\left\vert \mathbf{r}%
-\mathbf{r}^{\prime}\right\vert ^{2}}{\ell^{2}}\ \left(  \frac{e^{\lambda
\tau_{\mathrm{e}}}-1}{e^{\lambda t}-1}\right)  ^{2}\ \left\langle \int_{0}%
^{t}\mathrm{d}\overline{t}\int_{0}^{t}d\overline{t}^{\prime}\ e^{\lambda
\overline{t}+\lambda\overline{t}^{\prime}}\ \sin{\left[  2\theta(\overline
{t})\right]  }\ \sin{\left[  2\theta(\overline{t}^{\prime})\right]
}\right\rangle \ .
\end{equation}

As before, we assume that the different pieces are uncorrelated and the angles
$\theta_{i}$ uniformly distributed.\ Therefore $\left\langle \sin\left[
2\theta_{i}\right]  \sin\left[  2\theta_{j}\right]  \right\rangle =\delta
_{ij}/2$ and%

\begin{align}
\left\langle \left(  \Delta S_{s}-\Delta S_{s^{\prime}}\right)  ^{2}%
\right\rangle  &  \simeq\frac{\alpha^{2}}{2}\left(  \frac{m_{0}v^{2}}{\ell
}\right)  ^{2}\left\vert \mathbf{r}-\mathbf{r}^{\prime}\right\vert ^{2}\left(
\frac{e^{\lambda\tau_{\mathrm{e}}}-1}{e^{\lambda t}-1}\right)  ^{2}\sum
_{i=1}^{N_{s}}\left\langle \int_{t_{i-1}}^{t_{i}}\mathrm{d}\overline
{t}\ e^{\lambda\overline{t}}\right\rangle ^{2}\\
&  =\frac{\alpha^{2}}{2}\left(  \frac{m_{0}v^{2}}{\lambda\ell}\right)
^{2}\left\vert \mathbf{r}-\mathbf{r}^{\prime}\right\vert ^{2}\frac{\left(
e^{\lambda\tau_{\mathrm{e}}}-1\right)  ^{4}}{\left(  e^{\lambda t}-1\right)
^{2}}\frac{e^{2\lambda N_{s}\tau_{\mathrm{e}}}-1}{e^{2\lambda\tau_{\mathrm{e}%
}}-1}=A\ \left\vert \mathbf{r}-\mathbf{r}^{\prime}\right\vert ^{2}\ ,
\label{eq:difactions}%
\end{align}

\noindent where we have taken the limit $\lambda t\gg1$, and defined%

\begin{equation}
A=\frac{\alpha^{2}}{2}\left(  \frac{m_{0}v^{2}}{\lambda\ell}\right)  ^{2}%
\frac{\left(  e^{\lambda\tau_{\mathrm{e}}}-1\right)  ^{3}}{e^{\lambda
\tau_{\mathrm{e}}}+1} \ .
\end{equation}

Our result (\ref{eq:difactions}) is analogous to Eq.~(\ref{eq:deltmdelsp})
obtained in the case of a perturbation by a quenched disorder. Obviously, the
factor $A$ is different in both cases, but we use the same notation to stress
the similar role as just a prefactor of $M^d$. Performing again a Gaussian integral of
$M^{\mathrm{d}}$ over $\mathbf{r}-\mathbf{r}^{\prime}$ we obtain%

\begin{equation}
M^{\mathrm{d}}(t)=\left(  \frac{\sigma^{2}}{\pi\hbar^{2}}\right)  ^{d}%
\int\mathrm{d}\mathbf{r}\sum_{s}\ C_{s}^{2}\left(  \frac{2\pi\hbar^{2}}%
{A}\right)  ^{d/2}\exp{\left[  -\frac{2\sigma^{2}}{\hbar^{2}}\left(
{\overline{\mathbf{p}}}_{s}-\mathbf{p}_{0}\right)  ^{2}\right]  }\ .
\end{equation}

Under the same assumptions than in Sec.~\ref{subsec:leiaccs}, we are lead to a
result equivalent to that of Eq.~(\ref{eq:oversqdi3}),%

\begin{equation}
M^{\mathrm{d}}(t)\simeq\overline{A}e^{-\lambda t}, \label{eq:mdiaglor2}%
\end{equation}

\noindent with $\overline{A}=[\sigma m_{0}/(A^{1/2}t)]^{d}$. Therefore, for
long times the diagonal part of the Loschmidt echo decays with a rate given by
the classical Lyapunov exponent of the system,%

\begin{equation}
\lim_{t\rightarrow\infty}\left(  -\frac{1}{t}\ \ln\left[  M^{\mathrm{d}%
}(t)\right]  \right)  =\lambda. \label{eq-Lambda}%
\end{equation}
Of course this limit actually means $t\gg1/\lambda$ but still lower than the
time at which either localization or finite size effect appears. In the next
chapter we will study the competition between the diagonal and non-diagonal contributions.

\subsection{Diagonal vs. non-diagonal contributions}

As we have previously shown, the Loschmidt echo is made out of non-diagonal
and diagonal components, and within the time scales above specified, it can be
written as%

\begin{equation}
M(t)=\exp\left[  -\frac{vt}{\tilde{\ell}}\right]  +\overline{A}\exp\left[
-\lambda t\right]  \ .
\end{equation}

\noindent Such a result holds for the perturbation $\Sigma$ that we have
discussed in this section (Eq.~(\ref{eq-Perturbation})), as well as for the
quenched disorder of Sec.~\ref{sec:semi} (Eq.~(\ref{eq:irrevpart}). The only
difference lays in the form of the \textquotedblleft elastic mean free path"
$\tilde{\ell}$ and the prefactor $\overline{A}$, both of which are
perturbation dependent. The decay of the LE will be controlled by the slowest
of the two rates. A weak perturbation implies $\tilde{\ell}>v/\lambda$ and a
dominance of the non-diagonal term, while for sufficiently strong
perturbations verifying $\tilde{\ell}<v/\lambda$ (but weak enough in order not
to modify appreciably the classical trajectories), the diagonal term (governed
by the Lyapunov exponent) sets the decay of the LE. In Ref.
\onlinecite {cit-Jacquod01} the regime of dominance of the non-diagonal and
diagonal component has been respectively interpreted and referred to as a
Fermi Golden Rule Lyapunov regimes and we will use both terminologies in the
discussions that follow.

The Lyapunov regime is remarkable in the sense that its decay rate is an
intrinsic property of the system and does not depend on the perturbation
that gives rise to the decay. This behavior, predicted in
Ref.~\onlinecite{cit-Jalabert-Past} has been observed in numerical simulations
done on a number of systems \cite{cit-PhysicaA}$^{-}$\cite{cit-Mirlin}.

From the previous discussion it is clear that the Lyapunov regime can only be
observed beyond a critical value of the perturbation. The condition stated
above for the strength of the perturbation, along with
Eq.~(\ref{eq-FGRexponent}), yields for the model discussed in this section a
critical value of the perturbation parameter $\alpha$ beyond which the
Lyapunov regime is obtained,%

\begin{equation}
\alpha_{\mathrm{c}}=\frac{2\hbar}{m_{0}}\sqrt{\frac{\lambda}{v^3 \ell}}.
\label{eq-AlfaCritico}%
\end{equation}

\noindent We will discuss in Sec. IV the physical consequences of the above
critical value and its dependence on various physical parameters.

We finish this chapter with the discussion of the perturbation dependent
(non-diagonal) regime. In this Fermi Golden Rule regime the LE is equal to the
return probability $P(t)=\left\vert \left\langle \psi_{0}\right\vert
\exp\left[  -\mathrm{i}\left(  \mathcal{H}_{0}+\Sigma\right)  t/\hbar\right]
\left\vert \psi_{0}\right\rangle \right\vert ^{2}$, whose decay rate does not
show saturation at the Lyapunov exponent but rather follows Eq.
(\ref{eq-FGRexponent}) for the whole range of
parameters\cite{cit-WisniackiCohen}. The full connection between the exponent of
Eq.~(\ref{eq-FGRexponent}) and the one we would get from a complete FGR
approach, was clarified using a random matrix treatment
\cite{cit-SmoothBilliard}.
In fact, a rough estimation of the Fermi
golden rule is obtained considering a particle moving along the principal axis
$\overleftrightarrow{m}$ of a square box of sides $L_{x}=L_{y}=\ell$.
The available density of states $\frac{1}{\Delta}$, corresponds to a
1-d tube of length $\ell.$ Hence%
\begin{align*}
\frac{1}{\tilde{\tau}}  &  =\frac{2\pi}{\hbar}\left\vert \Sigma\right\vert
^{2}\frac{1}{\Delta}\\
&  \simeq\frac{2\pi}{\hbar}\left\vert \alpha\frac{p^{2}}{2m_{0}}\right\vert
^{2}\times\frac{\ell m_{0}^{{}}}{\pi\hbar p}\\
&  =m^{2}\frac{v^{3}}{4\hbar^{2}}\ell\alpha_{{}}^{2},
\end{align*}
in agreement with the semiclassical calculation of Eq. (\ref{eq-FGRexponent}).
Of course, this estimation does not make justice to the chaotic nature of
$\mathcal{H}_{0}$. This breaks the selection rules of our simple perturbation
and enables a random matrix approximation for $\Sigma$, that mixes
eigenstates of $\mathcal{H}_{0}$ which follow a Wigner-Dyson statistics. Hence
the perturbation breakdown of the FGR regime can not be calculated from the
parameters introduced above.

The LE has also been studied in systems where the perturbation structure
prevents the application of the Fermi Golden Rule. The result is that in
general the decay rate of the LE before the Lyapunov regime is given by the
width of the local density of states of the perturbation, which for particular
systems coincides with the exponent given by the FGR\cite{cit-Bunimovich}.

\section{Universality of the Lyapunov regime}

\label{sec:universality}

\subsection{Correspondence between semiclassical and numerical calculations}

\label{sec:numeric}

The semiclassical results obtained in the previous sections are valid in the
small wavelength limit, and relay on various uncontrolled approximations. It
is then important to perform numerical calculations for our model system in
order to compare against the semiclassical predictions, and to explore
parameter regimes inaccessible to the theory. In this section we use the same
numerical method of Ref.~\onlinecite {cit-Lorentzgas} for the Lorentz gas
(described in detail in appendix \ref{ape:numerics}), and extend the results
in order to sustain the discussion on the universality of the Lyapunov regime.
We will first focus on the behavior of the ensemble averaged Loschmidt echo,
followed by a thorough discussion of the averaging procedure and the individual 
behavior.

\begin{figure}[tb]
\centering \includegraphics*[width=8.3cm]{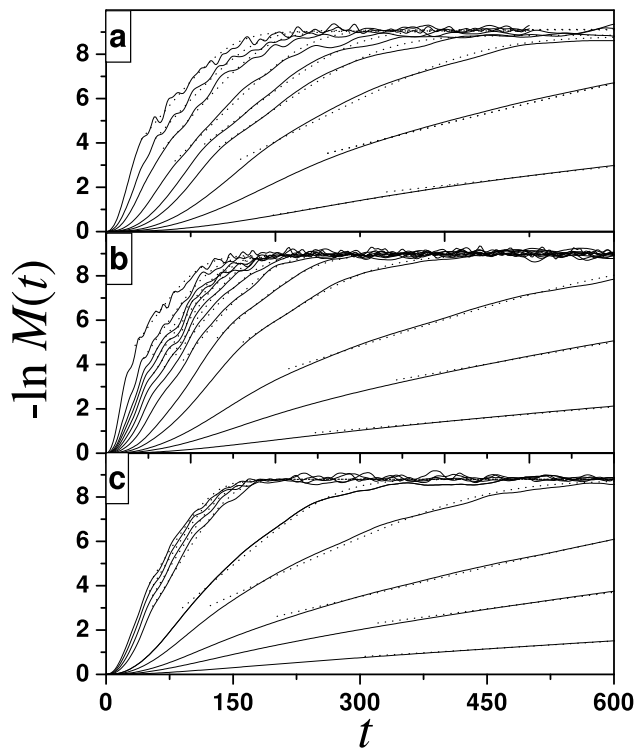}
\caption{Time decay of the Loschmidt echo $M(t)$ for different values of the
perturbation strength ($\alpha$) and concentration of impurities ($c$). (a)
$c=0.157$ and $\alpha=$0.004, 0.007, 0.01, 0.015, 0.02, 0.03, 0.05, 0.07, 0.1
(from top to bottom); (b) $c=0.195$ and $\alpha=$0.004, 0.007, 0.01, 0.015,
0.02, 0.03, 0.04, 0.05, 0.06, 0.07, 0.08, 0.1, 0.15; (c) $c=0.289$ and
$\alpha=$0.004, 0.007, 0.01, 0.015, 0.02, 0.03, 0.04, 0.05, 0.06, 0.07. The
time is measured in units of $\hbar/V$, where $V$ is the hopping term of the
tight-binding model (see appendix \ref{ape:numerics}). The doted lines
represent the best fits to the decay, as described in the text.}%
\label{fig-PanelesMvst}%
\end{figure}

We typically worked with disks of radius $R=20a$ , and with a Fermi wavelength
$\lambda_{F}=2\pi/k_{F}=16/3a$. Here, \ $a$ is the irrelevant lattice unit
of our tight-binding model, which is decreased until the results only depend
on the relation between physical parameters. The smallest system-size allowing
to observe the exponential decay of $M(t)$ over a large interval was found to
be $L=200a$ which means the consideration of a Hilbert space with 4$\times
$10$^{4}$ states. We calculated $M(t)$ for different strengths of the
perturbation $\alpha$ and concentration of disks $c$. In
Fig.~\ref{fig-PanelesMvst} we show our results for $c=0.157$, $0.195$ and
$0.289$ (panels $a$ to $c$ respectively), and different values of $\alpha$.

The time evolution of the LE presents various regimes. Firstly, for very short
times, $M(t)$ exhibits a Gaussian decay, $M(t)=\exp\left[  -b\alpha^{2}%
t^{2}\right]  $, where $b$ is a parameter that depends on the initial state,
the dynamics of $\mathcal{H}_{0}$ and the form of the perturbation $\Sigma$.
This initial decay corresponds to the overlap of the perturbed and unperturbed
wave-packets whose centers separate linearly with time by the sole effect of
the perturbation. This regime ends approximately at the typical time of the
first collision.

Secondly, for intermediate times we find the region of interest for the
semiclassical theory. In this time scale the LE decays exponentially with a
characteristic time $\tau_{\phi}$. We reserve the symbol $\tau_{\phi}$ for the
decay rate, in view of its interpretation in terms of quantum decoherence (as
we discuss in Sec.~\ref{sec:wigner}). For small perturbations, $\tau_{\phi}$
depends on $\alpha$. We observe that for all concentrations there is a
critical value $\alpha_{\mathrm{c}}$ beyond which $\tau_{\phi}$ is independent
of the perturbation. Clearly, the initial perturbation-dependent Gaussian
decay prevents the curves to be superimposed.

Finally, for very large times the LE saturates at a value $M_{\infty}$ that
depends on the system size $L$. This regime is discussed in detail in the next
chapters. However, let us observe that in the crossover between the
exponential decay and the long time saturation there is a power-law decay with
a perturbation independent exponent. This is a manifestation of the underlying
diffusive dynamics that leads to the isotropic state. Therefore, it could be
related to the Ruelle-Perricot resonances of the classical Perron-Frobenius
evolution operator used to calculate a classical version of the LE
\cite{cit-CasatiClassic}.

\begin{figure}[tb]
\centering \includegraphics*[width=8.3cm]{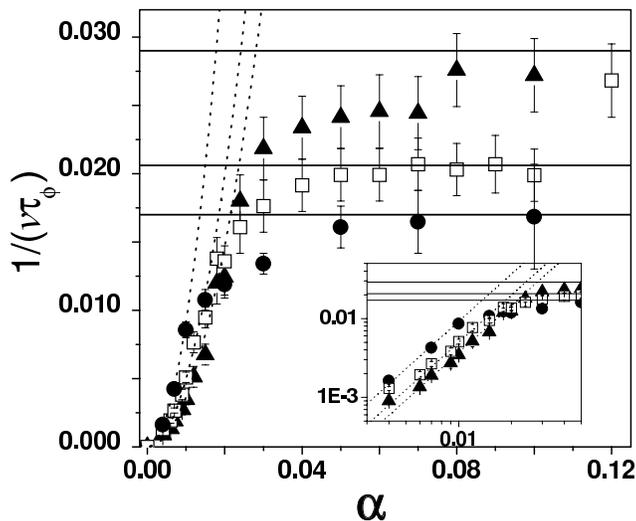}
\caption{Extracted values of the decay rate $1/\tau_{\phi}$ of the LE as a
function of the perturbation strength $\alpha$ for the three concentrations of
Fig.~\ref{fig-PanelesMvst}. The rates are normalized to the group velocity of
the initial wave-packet $\nu$ ($1/(\nu\tau_{\phi})$) is given in units of
$a^{-1}$; $c=0.157$ (circles), $0.195$ (squares) and $c=0.289$ (triangles).
The solid lines are the corresponding classical Lyapunov exponents and the
dashed lines are fits to the quadratic behavior predicted by
Eq.~(\ref{eq-FGRexponent}). The predicted coefficients for the three
concentrations are $72 a^{-1}$, $55 a^{-1}$ and $33 a^{-1}$, while the
obtained ones are $92 a^{-1}$, $50 a^{-1}$ and $37 a^{-1}$ respectively. In
the inset, a log-log scale of the same data to show the quadratic increase of
$1/\tau_{\phi}$ for small perturbations.}%
\label{fig-TauPhivsAlfa}%
\end{figure}

In order to compare our numerical results of $M(t)$ with the semiclassical
predictions, we extract $\tau_{\phi}$ by fitting $\ln M(t)$ to $\ln\left[
A\exp(-t/\tau_{\phi})+M_{\infty}\right]  .$ The dashed lines in
Fig.~\ref{fig-PanelesMvst} correspond to the best fits obtained with this
procedure. The values of $\tau_{\phi}$ for the different concentrations are
shown as a function of the perturbation strength in
Fig.~\ref{fig-TauPhivsAlfa}. In agreement with our analytical results of the
previous section, we see that $1/\tau_{\phi}$ grows quadratically with the
perturbation strength up to a critical value $\alpha_{\mathrm{c}}$, beyond
which a plateau appears at the corresponding Lyapunov exponent. The dashed
lines are the best fit to a quadratic behavior. The values obtained in this
way agree with those predicted by the semiclassical theory (Eq.
(\ref{eq-FGRexponent})) for the non-diagonal (FGR) term. The saturation values
above $\alpha_{\mathrm{c}}$ are well described by the corresponding Lyapunov
exponents (solid lines), in agreement with the semiclassical prediction
(Eq.~(\ref{eq-Lambda})). The very good quantitative agreement between the
semiclassical and numerical calculations for the Lorentz gas (as well as in
the case of other models
\cite{cit-Jacquod01,cit-Bunimovich,cit-SmoothBilliard}) strongly supports the
generality of the saturation of $\tau_{\phi}$ at a critical value of the
perturbation strength.

The FGR exponent, which depends on $\mathcal{H}_{0}$ but not much on its
chaoticity \cite{cit-JacquodInt}, is given by the typical squared matrix 
element of $\Sigma$, and the density of connected final states $1/\Delta$. Hence,
different $\mathcal{H}_{0}$ change the wave-functions. That is why we observe
that, for fixed perturbation strength $\alpha$, the factor $v/\tilde{\ell}$
depends on the concentration of impurities of $\mathcal{H}_{0}$ (see inset of
Fig.~\ref{fig-TauPhivsAlfa}, where a log-log scale has been chosen in order to
magnify the small perturbation region).

Notably, the dependence of $v/\tilde{\ell}$ with $\mathcal{H}_{0}$ leads to a
counter-intuitive effect (clearly observed in the inset of
Fig.~\ref{fig-TauPhivsAlfa}), namely that the critical value needed for the
saturation of $1/\tau_{\phi}$ is smaller for less chaotic systems (smaller
$\lambda$). The reason for this is that in more dilute systems $\Sigma$ is
constant over larger straight pieces of trajectories (in between collisions),
leading to a larger perturbation of the quantum phase and resulting in a
stronger effective perturbation.

\subsection{Universality of the Lyapunov regime in the semiclassical limit}

\label{sec:univ}

\begin{figure}[tb]
\centering \includegraphics*[width=8.3cm]{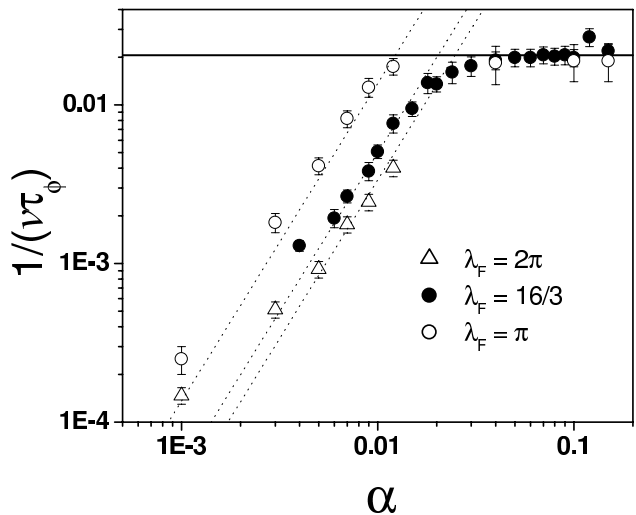}
\caption{Decay rates $1/\tau_{\phi}$ for different wavelengths $\lambda_{F}$
of the initial wave-packet for a concentration $c=0.195$ with the same units
as in Fig~\ref{fig-TauPhivsAlfa}. Solid line: classical Lyapunov exponent.
Dashed line: the FGR quadratic behavior. Note that for decreasing $\lambda
_{F}$ the critical perturbation diminishes, implying a collapse of the Fermi
Golden Rule regime.}%
\label{fig-AlfaCvsK}%
\end{figure}

Our semiclassical analysis yielded a critical value of the perturbation to
enter in the Lyapunov regime (Eq. (\ref{eq-AlfaCritico})), that vanishes in
the semiclassical limit, $\alpha_{\mathrm{c}}\rightarrow0$ for $\hbar$ (or
$\lambda_{F}$) $\rightarrow0$, implying the collapse of the Fermi Golden Rule
regime. This behavior is reproduced by our numerical calculations
(Fig.~\ref{fig-AlfaCvsK}). There, we decreased $\lambda_{F}$ while keeping
fixed the size $\sigma$ of the initial wave packet. A point that should not be
over-sighted is that the perturbation $\Sigma$ (Eq. (\ref{eq-Perturbation})),
for a given value of the parameter $\alpha$, scales with the energy in a way
that the underlying classical trajectories are always affected in the same way
by the perturbation. The extracted crossover values of $\alpha_{\mathrm{c}}$
are in quantitative agreement with Eq. (\ref{eq-AlfaCritico}), decreasing with
$\lambda_{F}$ in the interval that we were able to test.

Other choices of the perturbation $\Sigma$, such as the quenched disorder of
Refs.~\onlinecite{cit-Jalabert-Past} and \onlinecite{cit-SmoothBilliard}, can
be shown to give critical values that decrease with decreasing $\hbar$ as in
Eq.~(\ref{eq-AlfaCritico}), provided that the perturbation is scaled to the
proper semiclassical limit. That is, for a fixed perturbation potential, we
should take the limit of $\lambda_{F}\rightarrow0$. As a result, if we keep
$\hbar$ constant and decrease $\lambda_{F}$ by increasing the particle energy,
we should scale up the perturbation potential consistently (assuming that
$\mathcal{H}_{0}$ generates the same dynamics at all energies).

We conclude that, in the semiclassical limit, any perturbation will be strong
enough to put us in the Lyapunov regime, in consistency with the
hypersensitivity expected for a classical system. This is not unexpected as in
this limit the Ehrenfest time diverges and the correspondence principle should prevail.

\begin{figure}[tb]
\centering \includegraphics*[width=8.3cm]{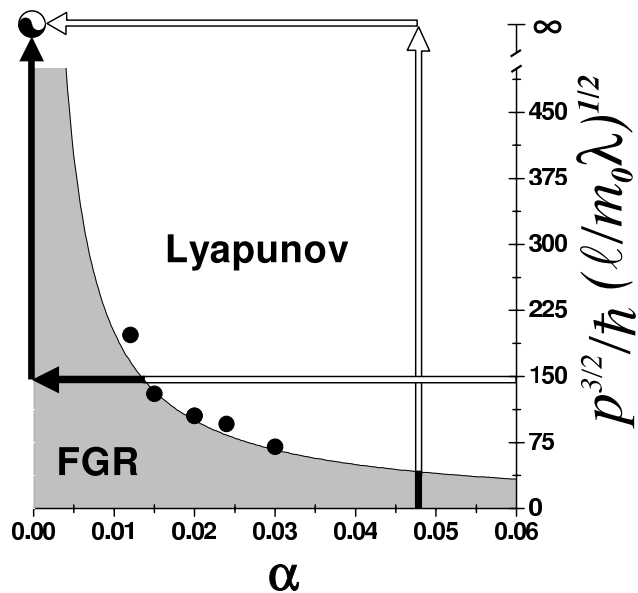} \vspace{0.5cm}%
\caption{Regime diagram for the Loschmidt echo as a function of the
perturbation and the energy (or inverse $\hbar$). The grayed area is the FGR
regime, while the clear one is the Lyapunov regime. The line that divides both
regimes is Eq. (\ref{eq-AlfaCritico}). The dots are the numerical values
obtained from Figs. \ref{fig-TauPhivsAlfa} and \ref{fig-AlfaCvsK}. The arrows
schematize the possible ordering of the classical double limit of the
perturbation and the wavelength going to zero. Notice how the lower one gives
always zero while the upper (correct) one gives $\lambda$ since it remains
always in the Lyapunov regime.}%
\label{fig-Diagramafases}%
\end{figure}

We can draw a the critical perturbation strength separating FGR from Lyapunov
regimes versus a scaling parameter determined by the particle energy (or
inverse $\hbar$), as shown in Fig.~\ref{fig-Diagramafases}. The shaded region
corresponds to the Fermi Golden Rule regime and the clear one to the Lyapunov
regime. The line that divides both phases is given by
Eq.~(\ref{eq-AlfaCritico}), and the dots correspond to numerical values of
$\alpha_{\mathrm{c}}$ extracted from Figs.~\ref{fig-TauPhivsAlfa} and
\ref{fig-AlfaCvsK}. Of course, there is another transition from FGR to
perturbation appearing when $\Sigma\simeq\Delta,$ which we avoid drawing since
$\alpha_{\Delta}\ll\alpha_{\mathrm{c}}$. This perturbative value also goes to
zero in the semiclassical limit of $\lambda_{F}\rightarrow0.$ Also the
Lyapunov regime is bounded from above by an $\hbar$ independent critical value
$\alpha_{\mathrm{p}}$ marking the classical breakdown that we discuss bellow.

The interesting conceptual feature highlighted by Fig.~\ref{fig-Diagramafases}%
, is the importance of the order in which we take the limits of $\Sigma$ and
$\lambda_{F}$ going to zero. Two distinct results are obtained for the
different order in which we can take this double limit. As depicted in the
figure (with arrows representing the limits), $\lim_{\lambda_{F}\rightarrow
0}\lim_{\Sigma\rightarrow0}1/\tau_{\phi}=0$. On the other hand, taking the
inverse (more physical) ordering $\lim_{\Sigma\rightarrow0}\lim_{\lambda
_{F}\rightarrow0}1/\tau_{\phi}=\lambda$ the semiclassical result is obtained.
The resulting \textquotedblleft phase diagram\textquotedblright%
\ representation for the different regimes of the LE serves us to remember
that most often one is working in the thermodynamic side corresponding to the
Lyapunov region.

Our semiclassical theory clearly fails when the perturbation is strong enough
(or the times long enough) to appreciably modify the classical trajectories.
This would give an upper limit (in perturbation strength) for the results of
Sec. III. A more stringent limitation comes from the finite value of $\hbar$,
due to the limitations of the diagonal approximations and linear expansions of
the action that we have relied on. In other systems, like the quenched
disorder in a smooth stadium \cite{cit-SmoothBilliard}, the upper critical
value of the perturbation (for exiting the Lyapunov regime) can be related to
the transport mean free path of the perturbation $\tilde{\ell}_{\mathrm{tr}}$,
which is defined as the length scale over which the classical trajectories are
affected by the disorder\cite{JMP}.

We can obtain in our system an estimate of $\tilde{\ell}_{\mathrm{tr}}$ by
considering the effect of the perturbation on a single scattering event. The
difference $\delta\theta$ between the perturbed and unperturbed exit angles
after the collision can be obtained using Eqs. (\ref{eq-ReflectionLaw}), which
results in
\begin{equation}
\delta\theta\sim4n_{x}n_{y}\left(  \frac{\mathbf{v}\cdot\mathbf{n}}{v}\right)
^{2} \ \alpha\ , \label{eq-DeltaAngulo}%
\end{equation}
where $\mathbf{v}$ is the initial velocity of the particle and $\mathbf{n}$ is
the normal to the surface.

Assuming that the movement of the particle is not affected by chaos
(non-dispersive collisions), one can do a random walk approach and
estimate the mean square distance after a time $\tau_{tr}$ from the
fluctuations of the angle in Eq.~(\ref{eq-DeltaAngulo}). We estimate
the transport mean free time as that at which the fluctuations are of the 
order of $R$, and obtain
\begin{equation}
\tilde{\ell}_{\mathrm{tr}}\simeq \frac{4 R^2}{3 \alpha^2\ell}\ ,
\label{eq-LdeTransporte}%
\end{equation}
assuming a uniform probability for the angle of the velocity as before.
Eq.~(\ref{eq-LdeTransporte}) is used to get the upper bound
perturbation $\alpha_{\mathrm{p}}$ for the end of the Lyapunov plateau,
\begin{equation}
\alpha_{\mathrm{p}}=\sqrt{\frac{4\lambda R^2}{3 \ell v}}\ . \label{alphap}%
\end{equation}
We obtain $\alpha_{\mathrm{p}}\simeq 0.23,$ $0.29$ and $0.43$ respectively 
for increasing
magnitude of the three concentrations shown in Fig.~\ref{fig-TauPhivsAlfa}.
It is rather
difficult to reach numerically these perturbations in our system, since the
initial Gaussian decay drives $M(t)$ very quickly towards its saturation
value, preventing the observation of an exponential regime. Despite this
difficulty, we observe in Fig.~\ref{fig-AlfaCvsK} that Lyapunov regime is a
plateau that ends up for sufficiently strong perturbations. For the range we
could explore the limiting values are in qualitative agreement with the
estimation from Eq.~(\ref{alphap}).

\subsection{Ehrenfest time and thermodynamic limit}

\label{sec:Ehrenfest}

We studied in the previous chapter the behavior of the Lyapunov regime in the
semiclassical limit $\hbar\rightarrow0$; let us now turn our attention to the
consequences of having a finite value of $\hbar$. In this case, one expects
the propagation of a quantum wave-packet to be described by the classical
equations of motion up to the Ehrenfest time $t_{E}$, after which the
quantum-classical correspondence breaks down \cite{cit-BermanZavslasky}.
Typically, $t_{E}$ is the time when interference effects become relevant, and
in a classically chaotic system it typically scales as $\ln[\hbar]$.

\begin{figure}[tb]
\centering \includegraphics*[width=8.3cm]{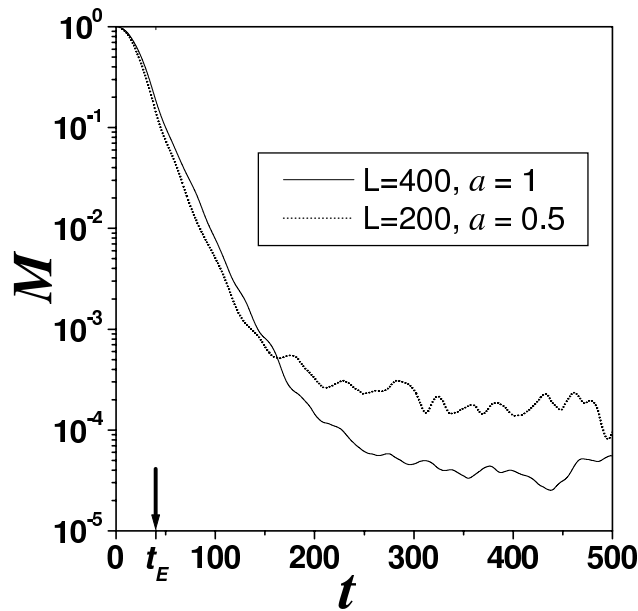}
\caption{$M(t)$ for two systems with the same number of states $N$ but
different sizes $L$, showing that the saturation time does not depend on $N$
but on the ratio $L/\sigma$. The Ehrenfest time is marked with an arrow.}%
\label{fig-EhrenfestTime}%
\end{figure}

In other systems where the Lyapunov regime of the LE has been observed, such
as chaotic maps or kicked systems, $t_{E}$ coincides with the saturation time
$t_{s}=1/\lambda$ $\ln[N]$. This is because in these systems the number of
states $N$ plays the role of an effective Planck's constant $\hbar
_{\mathrm{eff}}=1/N$. Therefore, when in these systems the LE is governed by a
classical quantity, the whole range of interest occurs before the Ehrenfest
time. This observation lead Benenti and Casati\cite{cit-BenentiCasati} to
propose that the independence of the decay rate on the perturbation strength
is a trivial consequence of the quantum-classical correspondence before
$t_{E}$.

In the Lorentz gas, however, we can differentiate between the time scales
$t_{s}$ and $t_{E}$ by appropriately controlling the parameters. The
saturation time is given by
\begin{equation}
t_{s}\simeq\frac{2}{\lambda}\ln\frac{L}{\sigma} \ .
\end{equation}

\noindent The Ehrenfest time, defined as the time it takes for a minimal
wave-packet of wavelength $\lambda_{F}$ to spread over a distance of the order
of $R$ \cite{cit-AleinerLarkin}, is given by
\begin{equation}
t_{E}\simeq\frac{1}{\lambda}\ln\frac{2R}{\lambda_{F}} \ .
\end{equation}

Our numerical calculations support these approximations. We show in
Fig.~\ref{fig-EhrenfestTime} $M(t)$ for two systems with the same number of
states $N$ (hence same $\hbar_{\mathrm{eff}}$), but two different sizes $L$.
$N$ is controlled by the discretization step $a=L/N$. The different saturation
values (and times) observed imply that for the Lorentz gas $t_{E}$ and $t_{s}$
are independent of each other. Clearly we can study the LE for times
arbitrarily larger than $t_{E}$ by increasing the system size $L,$ which
controls $t_{s},$ and keeping all other parameters (including $t_{E}$, marked
with a dashed line in Fig.~\ref{fig-EhrenfestTime}) fixed. We can see that the
exponential decay of the LE continues for times larger than $t_{E}$, up to the
saturation time.

\begin{figure}[tb]
\centering \includegraphics*[width=8.3cm]{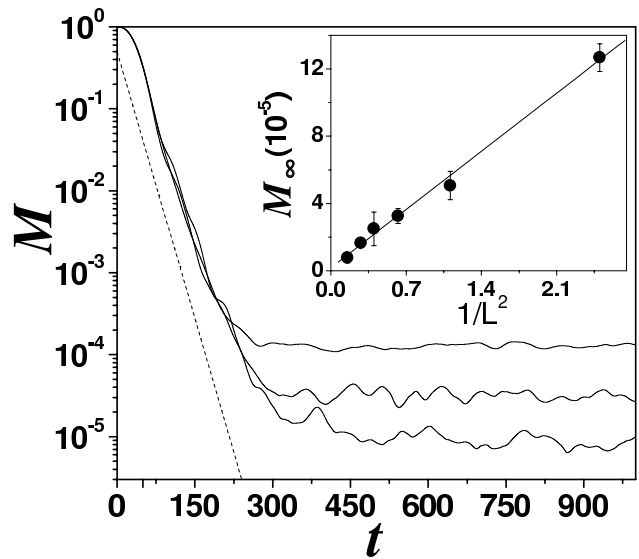} \vspace{0.5cm}%
\caption{$M(t)$ for different system sizes $L=200,400$ and $800$ showing the
longer exponential regime for larger $L$. Inset: saturation value $M_{\infty}$
as a function of inverse system size. The straight line is the fit $M_{\infty
}=(0.6)(\sigma/L)^{2}$.}%
\label{fig-MtVariacionL}%
\end{figure}

The above results are further evidence of the universality of the Lyapunov
regime, for it persists for arbitrarily large times in the thermodynamic limit
of the size of the system going to infinity. This is exemplified in
Fig.~\ref{fig-MtVariacionL}, where we show $M(t)$ for increasing sizes
($L=200a,400a$ and $800a$) for a fixed concentration ($c=0.195$) and
perturbation ($\alpha=0.024$). Notice how the exponential decay regime
extends, as $L$ grows, for times larger than $t_{E},$ where the correspondence
principle does not prevail. The survival of a classical signature of the
quantum dynamics after the Ehrenfest time is due to a more complex effect,
namely the decoherence that washes out terms of quantum nature. We will
discuss this process in detail in the next chapter.

In the inset of Fig.~\ref{fig-MtVariacionL} we see the saturation value
$M_{\infty}$ as a function of the inverse system size $1/L^{2}$. This
dependence was expected from earlier works on the LE \cite{cit-Peres}.
Supposing that for long times the chaotic nature of the system will equally
mix the $\tilde{N}=\left(  L/\sigma\right)  ^{2}$ levels appreciably
represented in the initial state with random phases $\phi_{j},$ we write
\begin{equation}
M_{\infty}=\frac{1}{\tilde{N}^{2}}\left|  \sum_{j}\exp\left[  \mathrm{i}%
(\phi_{j}-\phi_{j}^{\prime})\right]  \right|  ^{2}=\frac{1}{\tilde{N}} \ .
\end{equation}

We also show in a straight line the best fit to the data, $M_{\infty}%
=(0.6\pm0.1)\left(  \sigma/L\right)  ^{2}$ which confirms the prediction.

\subsection{Individual vs. ensemble-average behavior}

In order to make analytical progress, in our semiclassical calculations and in
those of Ref.~\onlinecite{cit-Jalabert-Past}, an ensemble average was
introduced (over realizations of the quenched disordered perturbation or over
initial conditions). This approximation raises the question of whether the
exponential decay of $M(t)$ is already present in individual realizations or,
on the contrary, the averaging procedure is a crucial ingredient in obtaining
a relaxation rate independent of the perturbation \cite{cit-Silvestrov}.

As it was discussed in Sects. \ref{sec:semi} and \ref{sec:lorentz}, for
trajectories longer than the correlation length $\xi$ of the perturbation, the
contributions to $\Delta S$ from segments separated by more than $\xi$ are
uncorrelated. This leads us to consider that the decay observed for a single
initial condition will be equivalent to that of the average. In this section
we test this idea numerically.

\begin{figure}[tb]
\centering \includegraphics*[width=8.3cm]{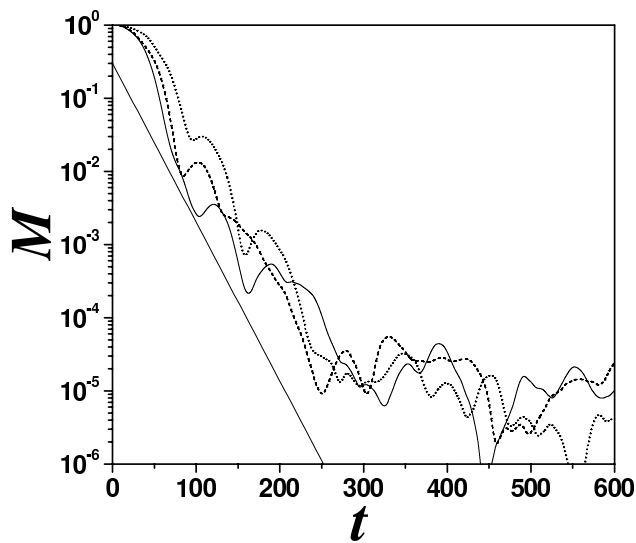}
\caption{$M(t)$ for three different single initial conditions of the
wave-packet. All the curves oscillate around the straight line, which is the
decay corresponding to the Lyapunov exponent.}%
\label{fig-CurvasIndividuales}%
\end{figure}

For large enough systems presenting a large saturation time, we expect $M(t)$
to fluctuate around an exponential decay. This expectation is clearly
supported by our numerical results shown in Fig.~\ref{fig-CurvasIndividuales},
where we present $M(t)$ for three different initial conditions in a system
with $L=800a$ and fixed $\alpha=0.024.$ An exponential decay with the
semiclassical exponent is shown for comparison (thin solid line).

In order to obtain the exponent of the decay with a good precision, we can
calculate $M(t)$ for a single initial condition in a large enough system.
Alternatively, our results show that it is correct to obtain the exponent
through an ensemble average to reduce the size of the fluctuations. However,
as the former method is computationally much more expensive, we resort to the latter.

This situation is analogous to the classical case where one obtains the
Lyapunov exponent from a single trajectory taking the limit of the initial
distance going to zero and the time going to infinity, or else resorts to more
practical methods like Benettin \emph{et al} algorithm that average distances
over short evolutions.

Notice that in the Lorentz gas the average over initial conditions and the
average over realizations of the impurities positions are equivalent. In all
cases we have implemented the last choice for being computationally
convenient, and we use the term initial conditions to refer also to
realizations of $\mathcal{H}_{0}$.

In particular for our calculations, the average is constrained to those
systems where the classical trajectory of the wave-packet collides with at
least one of the scatterers. This restriction helps avoiding those
configurations where a ``corridor'' exists, in which case $M(t)$ presents a
power-law decay possibly related to the behavior found in integrable systems
\cite{cit-JacquodInt}.

\subsection{Effect of the average procedure}

The averaging of quantities that fluctuate around an exponential decay is a
delicate matter, since different procedures can lead to quite different
results. In particular, for the LE it has been noted that averaging $M(t)$
over initial conditions can result in an exponential decay different than the
one of a single initial condition \cite{cit-WangLi,cit-Silvestrov}. This effect can be
attained, for instance, if we suppose that $M(t)$ for single conditions decays
exponentially with a fluctuating exponent $\lambda+\delta\lambda$, where
$\delta\lambda$ is randomly distributed with uniform probability between
$-\sigma_{\lambda}$ and $\sigma_{\lambda}$. Given the exponential dependence
of $M(t)$ in $\lambda$, the phase space fluctuations of the Lyapunov exponent
will induce a difference between the average $\ln M(t)$ and that of $M(t)$.
The former procedure is more appropriate in order to have averages of the
order of the typical values. On the other hand, if the fluctuations of the
exponent are small, both procedures give similar results. This is the
situation we found in our model system.

\begin{figure}[tb]
\centering \includegraphics*[width=8.3cm]{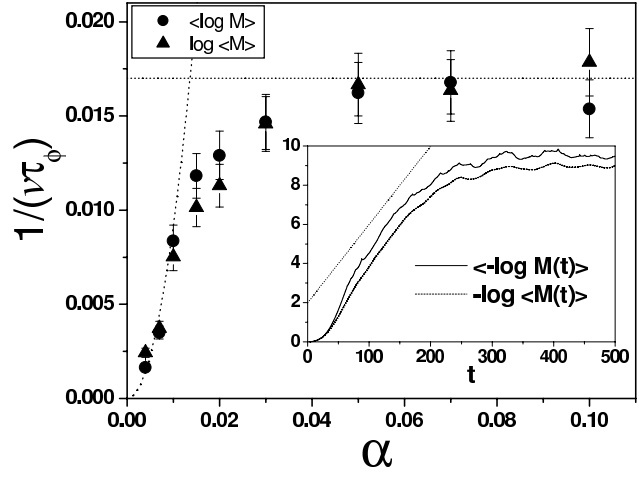}
\caption{Values of $\tau_{\phi}$ extracted from averaging the echo
$\left\langle M\right\rangle $ (triangles) and the logarithm of the echo
$\left\langle \log M \right\rangle $ (circles) for $c=0.289$ and $L=200$, as a
function of the perturbation strength. Both averaging procedures yield
approximate the same decay rate. Inset: typical curves of $\left\langle
M\right\rangle $ and $\left\langle \log M \right\rangle $ for the same set of
parameters. The straight dotted line represents an exponential decay with a
rate given by the semiclassical prediction.}%
\label{fig-MvsLogM}%
\end{figure}

For the Lorentz gas we calculated $\left\langle M(t)\right\rangle $ and
$\left\langle \ln M(t)\right\rangle $ and extracted the decay rates of the
exponential regime using the fit described in Sec.~\ref{sec:numeric}. Typical
results are shown in Fig.~\ref{fig-MvsLogM} as a function of the perturbation.
We observe that both averaging procedures give values of $\tau_{\phi}$ that
are indistinguishable from each other within the statistical error.

A typical set of curves of $M(t)$ averaged following the two procedures is
shown in the inset of Fig.~\ref{fig-MvsLogM}. We can see that, even though for
the short and long time regimes the two procedures give slightly different
results, the intermediate exponential decay regime has approximately the same
decay rate in both cases.

\section{Analysis of decoherence through the Loschmidt echo}

\label{sec:wigner}

\subsection{Classical evolution of the Wigner function}

As discussed in the introduction, the Loschmidt echo can be obtained from the
evolution of the Wigner function with the perturbed and unperturbed
Hamiltonians (Eq.~(\ref{eq-LEWigner})). Such a framework is particularly
useful in the study of decoherence, as the Wigner function is a privileged
tool to understand the connection between quantum and classical dynamics
\cite{bookBrack,cit-ZurekNature}.

The evolution of the wave-functions in terms of the propagators
(Eq.~(\ref{eq:defprop})) can be used to express the time-dependence of the
Wigner function as%

\[
W(\mathbf{r},\mathbf{p};t)=\frac{1}{(2\pi\hbar)^{d}}\int\mathrm{d}%
\delta\mathbf{r}\int\mathrm{d}\overline{\mathbf{q}}\int\mathrm{d}%
{\overline{\mathbf{q}}}^{\prime}\exp{\left[  \frac{\mathrm{i}}{\hbar
}\ \mathbf{p}\cdot\delta\mathbf{r}\right]  }\ K\left(  \mathbf{r}-\frac
{\delta\mathbf{r}}{2},\overline{\mathbf{q}};t\right)  \ K^{\ast}\left(
\mathbf{r}+\frac{\delta\mathbf{r}}{2},{\overline{\mathbf{q}}}^{\prime
};t\right)  \ \psi(\overline{\mathbf{q}},0)\ \psi^{\ast}({\overline
{\mathbf{q}}}^{\prime},0).
\]

\noindent Noting $W(\overline{\mathbf{r}},{\overline{\mathbf{p}}};0)$ the
initial Wigner function, we can write
\begin{align}
W(\mathbf{r},\mathbf{p};t)  &  =\frac{1}{(2\pi\hbar)^{d}}\int\mathrm{d}%
\delta\mathbf{r}\int\mathrm{d}\overline{\mathbf{r}}\int\mathrm{d}%
\delta{\overline{\mathbf{r}}}\int\mathrm{d}{\overline{\mathbf{p}}%
}\ W(\overline{\mathbf{r}},{\overline{\mathbf{p}}};0)\ \exp\left[
\frac{\mathrm{i}}{\hbar}(\mathbf{p}\cdot\delta\mathbf{r}-{\overline
{\mathbf{p}}}\cdot\delta{\overline{\mathbf{r}}})\right] \nonumber\\
\displaystyle  &  \times\ K\left(  \mathbf{r}-\frac{\delta\mathbf{r}}%
{2},\overline{\mathbf{r}}-\frac{\delta{\overline{\mathbf{r}}}}{2};t\right)
\ K^{\ast}\left(  \mathbf{r}+\frac{\delta\mathbf{r}}{2},\overline{\mathbf{r}%
}+\frac{\delta{\overline{\mathbf{r}}}}{2};t\right)  \ . \label{eq:evwig}%
\end{align}

\begin{figure}[tb]
\centering \includegraphics*[width=8.3cm]{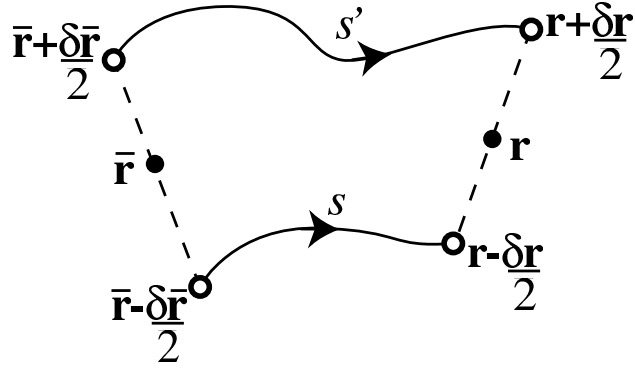}
\caption{Schematics of the classical trajectories involved in the
semiclassical approximation to the propagation of a Wigner function.}%
\label{1Wigner}%
\end{figure}

The semiclassical expansion of the propagators
(Eq.~(\ref{eq-SemiclasicalPropagator})) leads to the propagation of the Wigner
function by ``chords" \cite{Ozorio,Caio,Horacio92}, where pairs of trajectories
$(s,s^{\prime})$ traveling from $(\overline{\mathbf{r}}-\delta{\overline
{\mathbf{r}}}/2,\overline{\mathbf{r}}+\delta{\overline{\mathbf{r}}}/2)$ to
$(\mathbf{r}-\delta\mathbf{r}/2,\mathbf{r}+\delta\mathbf{r}/2)$ have to be
considered (Fig.~\ref{1Wigner}). In the leading order in $\hbar$ we can
approximate the above propagators by sums over trajectories going from
$\overline{\mathbf{r}}$ to $\mathbf{r}$%

\begin{subequations}
\label{allPRO}%
\begin{align}
K\left(  \mathbf{r}-\frac{\delta\mathbf{r}}{2},\overline{\mathbf{r}}%
-\frac{\delta{\overline{\mathbf{r}}}}{2};t\right)   &  =\sum_{s(\overline
{\mathbf{r}},\mathbf{r},t)}\ K_{s}(\mathbf{r},\overline{\mathbf{r}}%
;t)\ \exp\left[  \frac{\mathrm{i}}{2\hbar}\left(  {\overline{\mathbf{p}}}%
_{s}\cdot\delta{\overline{\mathbf{r}}}-\mathbf{p}_{s}\cdot\delta
\mathbf{r}\right)  \right]  \ ,\label{eq:RPO0}\\
\displaystyle K\left(  \mathbf{r}+\frac{\delta\mathbf{r}}{2},\overline
{\mathbf{r}}+\frac{\delta{\overline{\mathbf{r}}}}{2};t\right)   &
=\sum_{s^{\prime}(\overline{\mathbf{r}},\mathbf{r},t)}\ K_{s^{\prime}%
}(\mathbf{r},\overline{\mathbf{r}};t)\ \exp\left[  \frac{\mathrm{i}}{2\hbar
}\left(  -{\overline{\mathbf{p}}}_{s^{\prime}}\cdot\delta{\overline
{\mathbf{r}}}+\mathbf{p}_{s^{\prime}}\cdot\delta\mathbf{r}\right)  \right]
\ , \label{eq:PRO1}%
\end{align}

\noindent where ${\overline{\mathbf{p}}}_{s}$ ($\mathbf{p}_{s}$) and
${\overline{\mathbf{p}}}_{s^{\prime}}$ ($\mathbf{p}_{s^{\prime}}$) are the
initial (final) momenta of the trajectories $s$ and $s^{\prime}$,
respectively. The semiclassical evolution of the Wigner function is given by%

\end{subequations}
\[
W(\mathbf{r},\mathbf{p};t)=(2\pi\hbar)^{d}\int\mathrm{d}\overline{\mathbf{r}%
}\int\mathrm{d}{\overline{\mathbf{p}}}\ W(\overline{\mathbf{r}},{\overline
{\mathbf{p}}};0)\ \sum_{s,s^{\prime}}\ \delta\left(  {\overline{\mathbf{p}}%
}-\frac{{\overline{\mathbf{p}}}_{s}+{\overline{\mathbf{p}}}_{s^{\prime}}}%
{2}\right)  \ \delta\left(  \mathbf{p}-\frac{\mathbf{p}_{s}+\mathbf{p}%
_{s^{\prime}}}{2}\right)  K_{s}\left(  \mathbf{r},\overline{\mathbf{r}%
};t\right)  \ K_{s^{\prime}}^{\ast}\left(  \mathbf{r},\overline{\mathbf{r}%
};t\right)  \ .
\]

The dominant contribution arises from the diagonal term $s=s^{\prime}$%

\begin{equation}
W_{\mathrm{c}}(\mathbf{r},\mathbf{p},t)=\int\mathrm{d}\overline{\mathbf{r}%
}\sum_{s(\overline{\mathbf{r}},\mathbf{r},t)}\ C_{s}\ \delta\left(
\mathbf{p}-\mathbf{p}_{s}\right)  \ W(\overline{\mathbf{r}},{\overline
{\mathbf{p}}}_{s};0)\ . \label{eq:evwigclas}%
\end{equation}

Using the fact that $C_{s}$ is the Jacobian of the transformation from
$\overline{\mathbf{r}}$ to $\mathbf{p}_{s}$, we have%

\begin{equation}
W_{\mathrm{c}}(\mathbf{r},\mathbf{p};t)=\int\mathrm{d}\mathbf{p}_{s}%
\ \delta\left(  \mathbf{p}-\mathbf{p}_{s}\right)  \ W(\overline{\mathbf{r}%
},{\overline{\mathbf{p}}}_{s};0)\ , \label{eq:evwigclas2}%
\end{equation}

\noindent where the trajectories considered now are those that arrive to
$\mathbf{r}$ with momentum $\mathbf{p}$. We note $(\overline{\mathbf{r}%
},{\overline{\mathbf{p}}})$ the pre-image of $(\mathbf{r},\mathbf{p})$ by the
equations of motion acting on a time $t$. That is, $(\mathbf{r},\mathbf{p}) =
X_{t} (\overline{\mathbf{r}},{\overline{\mathbf{p}}})$. The momentum integral
is trivial, and we obtain the obvious result%

\begin{equation}
W_{\mathrm{c}}(\mathbf{r},\mathbf{p};t)= W(\overline{\mathbf{r}}%
,{\overline{\mathbf{p}}};0) \ , \label{eq:evwigclas3}%
\end{equation}

\noindent with $(\overline{\mathbf{r}},{\overline{\mathbf{p}}}) = X_{t}%
^{-1}(\mathbf{r},\mathbf{p})$. Since $X_{t}$ conserves the volume in
phase-space, at the classical level the Wigner function evolves by simply
following the classical flow.

\subsection{Fine structure of the Wigner function and non-classical
contributions to the Loschmidt echo}

As indicated in Eq.~(\ref{eq-LEWigner}), the Loschmidt echo is given by the
phase-space trace of two Wigner functions associated with slightly different
Hamiltonians ($\mathcal{H}_{0}$ and $\mathcal{H}_{0}+\Sigma$). In order to
facilitate the discussion, we introduce the density (or partial trace)
$f_{\Sigma}$ writing the LE as%

\begin{equation}
M(t)=\int\mathrm{d}\mathbf{r}\ f_{\Sigma}(\mathbf{r},t)\ ,
\label{eq:deffsigma}%
\end{equation}

\noindent with%

\begin{align}
f_{\Sigma}(\mathbf{r},t)  &  =\frac{1}{(2\pi\hbar)^{d}}\ \int\mathrm{d}%
\mathbf{p}\int\mathrm{d}\delta\mathbf{r}\int\mathrm{d}\overline{\mathbf{r}%
}\int\mathrm{d}\delta{\overline{\mathbf{r}}}\int\mathrm{d}{\overline
{\mathbf{p}}}\int\mathrm{d}\delta\mathbf{r}^{\prime}\int\mathrm{d}%
{\overline{\mathbf{r}}}^{\prime}\int\mathrm{d}\delta{\overline{\mathbf{r}}%
}^{\prime}\int\mathrm{d}{\overline{\mathbf{p}}}^{\prime}\ W_{\Sigma}%
(\overline{\mathbf{r}},{\overline{\mathbf{p}}};0)\ W_{0}^{\ast}({\overline
{\mathbf{r}}}^{\prime},{\overline{\mathbf{p}}}^{\prime};0)\nonumber\\
\displaystyle  &  \times\ \exp{\left[  \frac{\mathrm{i}}{\hbar}\left(
\mathbf{p}\cdot\delta\mathbf{r}-{\overline{\mathbf{p}}}\cdot\delta
{\overline{\mathbf{r}}}\right)  \right]  }\exp{\left[  -\frac{\mathrm{i}%
}{\hbar}\left(  \mathbf{p}\cdot\delta\mathbf{r}^{\prime}-{\overline
{\mathbf{p}}}^{\prime}\cdot\delta{\overline{\mathbf{r}}}^{\prime}\right)
\right]  }\ K\left(  \mathbf{r}-\frac{\delta\mathbf{r}}{2},\overline
{\mathbf{r}}-\frac{\delta{\overline{\mathbf{r}}}}{2};t\right) \nonumber\\
\displaystyle  &  \times\ K^{\ast}\left(  \mathbf{r}+\frac{\delta\mathbf{r}%
}{2},\overline{\mathbf{r}}+\frac{\delta{\overline{\mathbf{r}}}}{2};t\right)
\ K^{\ast}\left(  \mathbf{r}-\frac{\delta\mathbf{r}^{\prime}}{2}%
,{\overline{\mathbf{r}}}^{\prime}-\frac{\delta{\overline{\mathbf{r}}}^{\prime
}}{2};t\right)  \ K\left(  \mathbf{r}+\frac{\delta\mathbf{r}^{\prime}}%
{2},{\overline{\mathbf{r}}}^{\prime}+\frac{\delta{\overline{\mathbf{r}}%
}^{\prime}}{2};t\right)  \ . \label{eq:fsigma}%
\end{align}

\begin{figure}[tb]
\centering \includegraphics*[width=8.3cm]{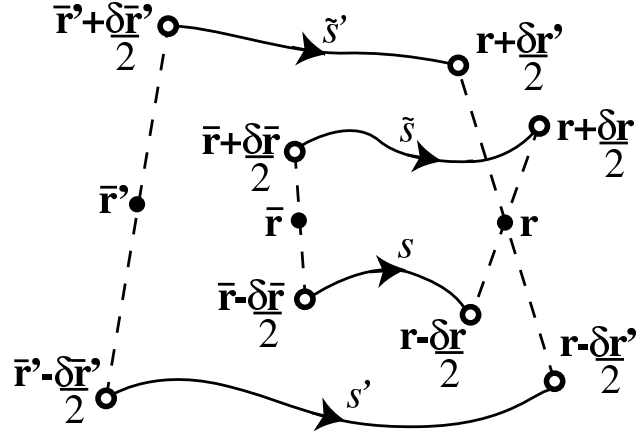}
\caption{ Four classical trajectories used to compute semiclassically the
Loschmidt echo through the evolution of two Wigner functions associated with
different Hamiltonians.}%
\label{2Wigners}%
\end{figure}

The semiclassical evolution of $f_{\Sigma}$ is given by four trajectories, as
illustrated in Fig.~\ref{2Wigners}.

As we have consistently done in this work, we take Gaussian wave-packet (of
width $\sigma$) as initial state. Its associated Wigner function reads%

\begin{equation}
W(\overline{\mathbf{r}},{\overline{\mathbf{p}}};0) = \frac{1}{(\pi\hbar)^{d}}
\ \exp{\left[  -\frac{(\overline{\mathbf{r}}-\mathbf{r}_{0})^{2}}{\sigma^{2}}
- \frac{({\overline{\mathbf{p}}}-\mathbf{p}_{0})^{2}\sigma^{2}}{\hbar^{2}}
\right]  } \ . \label{eq-WignerInicial}%
\end{equation}

Assuming that $\Sigma$ constitutes a small perturbation, after a few trivial
integrations we obtain%

\begin{align}
f_{\Sigma}(\mathbf{r},t)  &  =\frac{\sigma^{2}}{(2\pi^{3}\hbar^{4})^{d/2}}%
\int\mathrm{d}\delta\mathbf{r}\int\mathrm{d}\overline{\mathbf{r}}%
\int\mathrm{d}\delta{\overline{\mathbf{r}}}\int\mathrm{d}{\overline
{\mathbf{p}}}\int\mathrm{d}\delta{\overline{\mathbf{r}}}^{\prime}%
\int\mathrm{d}{\overline{\mathbf{p}}}^{\prime}\ \exp{\left[  \frac{\mathrm{i}%
}{\hbar}\left(  {\overline{\mathbf{p}}}^{\prime}\cdot\delta{\overline
{\mathbf{r}}}^{\prime}-{\overline{\mathbf{p}}}\cdot\delta{\overline
{\mathbf{r}}}\right)  \right]  }\ \exp{\left[  -\frac{2}{\sigma^{2}}%
(\overline{\mathbf{r}}-\mathbf{r}_{0})^{2}\right]  }\nonumber\\
\displaystyle  &  \times\ \exp{\left[  -\frac{\sigma^{2}}{\hbar^{2}}\left(
({\overline{\mathbf{p}}}-\mathbf{p}_{0})^{2}+({\overline{\mathbf{p}}}^{\prime
}-\mathbf{p}_{0})^{2}\right)  \right]  }\ \sum_{s,s^{\prime}}\ \sum
_{{\tilde{s}},{\tilde{s}}^{\prime}}\ \exp{\left[  -\frac{\mathcal{P}^{2}%
\sigma^{2}}{8\hbar^{2}}\right]  }\ K_{s}\left(  \mathbf{r}-\frac
{\delta\mathbf{r}}{2},\overline{\mathbf{r}}-\frac{\delta{\overline{\mathbf{r}%
}}}{2};t\right) \nonumber\\
\displaystyle  &  \times\ K_{s^{\prime}}^{\ast}\left(  \mathbf{r}-\frac
{\delta\mathbf{r}}{2},\overline{\mathbf{r}}-\frac{\delta{\overline{\mathbf{r}%
}}^{\prime}}{2};t\right)  \ K_{{\tilde{s}}}^{\ast}\left(  \mathbf{r}%
+\frac{\delta\mathbf{r}}{2},\overline{\mathbf{r}}+\frac{\delta{\overline
{\mathbf{r}}}}{2};t\right)  \ K_{{\tilde{s}}^{\prime}}\left(  \mathbf{r}%
+\frac{\delta\mathbf{r}}{2},\overline{\mathbf{r}}+\frac{\delta{\overline
{\mathbf{r}}}^{\prime}}{2};t\right)  \ . \label{eq:fevosemicla0}%
\end{align}

\noindent Where we have defined%

\begin{equation}
\mathcal{P} = {\overline{\mathbf{p}}}_{s}+{\overline{\mathbf{p}}}_{s^{\prime}%
}-{\overline{\mathbf{p}}}_{{\tilde s}}-{\overline{\mathbf{p}}}_{{\tilde
s}^{\prime}} \ .
\end{equation}

\noindent Now the trajectories $s$ and $s^{\prime}$ (${\tilde s}$ and ${\tilde
s}^{\prime}$) arrive to the same final point $\overline{\mathbf{r}}%
-\delta{\overline{\mathbf{r}}}/2$ ($\mathbf{r}+\delta\mathbf{r}/2$). Since the
initial wave-packet is concentrated around $\mathbf{r}_{0}$, we can further
simplify and work with trajectories $s$ and $s^{\prime}$ (${\tilde s}$ and
${\tilde s}^{\prime}$) that have the same extreme points. Therefore, we have%

\begin{align}
f_{\Sigma}(\mathbf{r},t)  &  =\frac{\sigma^{2}}{(2\pi^{3}\hbar^{4})^{d/2}%
}\ \int\mathrm{d}\delta\mathbf{r}\int\mathrm{d}\overline{\mathbf{r}}%
\int\mathrm{d}\delta{\overline{\mathbf{r}}}\ \exp{\left[  -\frac{2}{\sigma
^{2}}(\overline{\mathbf{r}}-\mathbf{r}_{0})^{2}\right]  }\ \sum_{s,s^{\prime}%
}\ \sum_{{\tilde{s}},{\tilde{s}}^{\prime}}\ \exp{\left[  -\frac{\mathcal{P}%
^{2}\sigma^{2}}{8\hbar^{2}}-\frac{2\sigma^{2}}{\hbar^{2}}\left(
\frac{\mathcal{R}}{4}-\mathbf{p}_{0}\right)  ^{2}-\frac{\delta{\overline
{\mathbf{r}}}^{2}}{2\sigma^{2}}\right]  }\nonumber\\
\displaystyle  &  K_{s}\left(  \mathbf{r}-\frac{\delta\mathbf{r}}{2}%
,\overline{\mathbf{r}}-\frac{\delta{\overline{\mathbf{r}}}}{2};t\right)
\ K_{s^{\prime}}^{\ast}\left(  \mathbf{r}-\frac{\delta\mathbf{r}}{2}%
,\delta{\overline{\mathbf{r}}}-\frac{\delta{\overline{\mathbf{r}}}}%
{2};t\right)  \ K_{{\tilde{s}}}^{\ast}\left(  \mathbf{r}+\frac{\delta
\mathbf{r}}{2},\overline{\mathbf{r}}+\frac{\delta{\overline{\mathbf{r}}}}%
{2};t\right)  \ K_{{\tilde{s}}^{\prime}}\left(  \mathbf{r}+\frac
{\delta\mathbf{r}}{2},\delta{\overline{\mathbf{r}}}+\frac{\delta
{\overline{\mathbf{r}}}}{2};t\right)  \ , \label{eq:fevosemicla}%
\end{align}

\noindent with%

\begin{equation}
\mathcal{R} = {\overline{\mathbf{p}}}_{s}+{\overline{\mathbf{p}}}_{s^{\prime}%
}+{\overline{\mathbf{p}}}_{{\tilde s}}+{\overline{\mathbf{p}}}_{{\tilde
s}^{\prime}} \ .
\end{equation}

\begin{figure}[tb]
\centering \includegraphics*[width=8.3cm]{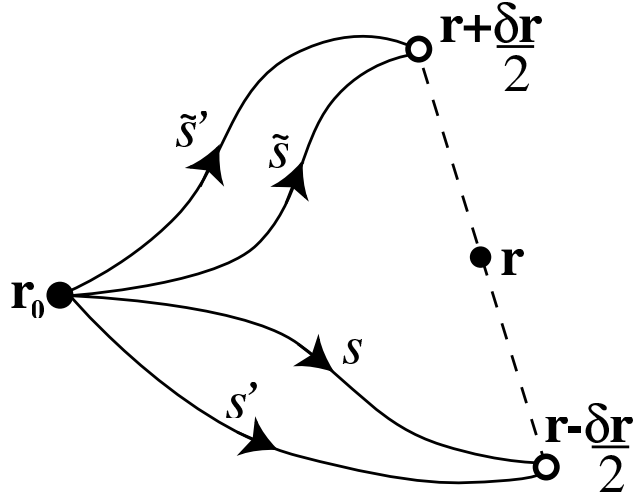}
\caption{ For fairly localized initial wave-packet, the four classical
trajectories contributing to the LE can be reduced to those starting at its
center $\mathbf{r}_{0}$. }%
\label{2WwSIP}%
\end{figure}

By the same considerations as before, we can reduce all four trajectories to
start at the center $\mathbf{r}_{0}$ of the initial wave-packet
(Fig.~\ref{2WwSIP})

\begin{align}
f_{\Sigma}(\mathbf{r},t)  &  =(4\pi\sigma^{2})^{d}\ \int\mathrm{d}%
\delta\mathbf{r}\ \sum_{s,s^{\prime}}\ \sum_{{\tilde{s}},{\tilde{s}}^{\prime}%
}\ \exp{\left[  -\frac{(\mathcal{P}^{2}+\mathcal{S}^{2}+\mathcal{T}^{2}%
)\sigma^{2}}{8\hbar^{2}}\right]  }\exp{\left[  -\frac{2\sigma^{2}}{\hbar^{2}%
}\left(  \frac{\mathcal{R}}{4}-\mathbf{p}_{0}\right)  ^{2}\right]
}\nonumber\\
\displaystyle  &  \times\ K_{s}\left(  \mathbf{r}-\frac{\delta\mathbf{r}}%
{2},\mathbf{r}_{0};t\right)  \ K_{s^{\prime}}^{\ast}\left(  \mathbf{r}%
-\frac{\delta\mathbf{r}}{2},\mathbf{r}_{0};t\right)  \ K_{{\tilde{s}}}^{\ast
}\left(  \mathbf{r}+\frac{\delta\mathbf{r}}{2},\mathbf{r}_{0};t\right)
\ K_{{\tilde{s}}^{\prime}}\left(  \mathbf{r}+\frac{\delta\mathbf{r}}%
{2},\mathbf{r}_{0};t\right)  \ , \label{eq:fevosemicla2}%
\end{align}

\noindent with%

\begin{subequations}
\label{allDEFST}%
\begin{align}
\mathcal{S}  &  = {\overline{\mathbf{p}}}_{s}-{\overline{\mathbf{p}}%
}_{s^{\prime}}+{\overline{\mathbf{p}}}_{{\tilde s}}-{\overline{\mathbf{p}}%
}_{{\tilde s}^{\prime}} \ ,\label{eq:DEFST0}\\
\displaystyle \mathcal{T}  &  = {\overline{\mathbf{p}}}_{s}+{\overline
{\mathbf{p}}}_{s^{\prime}}-{\overline{\mathbf{p}}}_{{\tilde s}}-{\overline
{\mathbf{p}}}_{{\tilde s}^{\prime}} \ . \label{eq:DEFST1}%
\end{align}

Given that%

\end{subequations}
\begin{equation}
\mathcal{P}^{2}+\mathcal{S}^{2}+\mathcal{T}^{2} = \left(  {\overline
{\mathbf{p}}}_{s}-{\overline{\mathbf{p}}}_{s^{\prime}}\right)  ^{2}+ \left(
{\overline{\mathbf{p}}}_{s}-{\overline{\mathbf{p}}}_{{\tilde s}}\right)  ^{2}+
\left(  {\overline{\mathbf{p}}}_{s}-{\overline{\mathbf{p}}}_{{\tilde
s}^{\prime}}\right)  ^{2}+ \left(  {\overline{\mathbf{p}}}_{s^{\prime}%
}-{\overline{\mathbf{p}}}_{{\tilde s}}\right)  ^{2}+ \left(  {\overline
{\mathbf{p}}}_{s^{\prime}}-{\overline{\mathbf{p}}}_{{\tilde s}^{\prime}%
}\right)  ^{2}+ \left(  {\overline{\mathbf{p}}}_{{\tilde s}}-{\overline
{\mathbf{p}}}_{{\tilde s}^{\prime}}\right)  ^{2} \ , \label{eq:summomsq}%
\end{equation}

\noindent and since the pairs of trajectories $(s,s^{\prime})$ and $({\tilde
s},{\tilde s}^{\prime})$ have the same extreme points, the dominant
contribution to $f_{\Sigma}$ will come from the terms with $s=s^{\prime}$ and
${\tilde s}={\tilde s}^{\prime}$. Such an identification minimizes the
oscillatory phases of the propagators, and corresponds to the first diagonal
approximation of the calculation of Sec.~\ref{sec:semi} and
Ref.~\onlinecite{cit-Jalabert-Past}. Within such an approximation we have%

\begin{align}
f_{\Sigma}(\mathbf{r},t)  &  =\left(  \frac{\sigma^{2}}{\pi\hbar^{2}}\right)
^{d}\ \int\mathrm{d}\delta\mathbf{r}\ \sum_{s,{\tilde{s}}}\ C_{s}%
\ C_{{\tilde{s}}}\ \exp{\left[  -\frac{\left(  {\overline{\mathbf{p}}}%
_{s}-{\overline{\mathbf{p}}}_{{\tilde{s}}}\right)  ^{2}\sigma^{2}}{2\hbar^{2}%
}-\frac{2\sigma^{2}}{\hbar^{2}}\left(  \frac{{\overline{\mathbf{p}}}%
_{s}+{\overline{\mathbf{p}}}_{{\tilde{s}}}}{2}-\mathbf{p}_{0}\right)
^{2}\right]  }\nonumber\\
\displaystyle  &  \exp{\left[  \frac{\mathrm{i}}{\hbar}\left(  \Delta
S_{s}\left(  \mathbf{r}-\frac{\delta\mathbf{r}}{2},\mathbf{r}_{0},t\right)
-\Delta S_{{\tilde{s}}}\left(  \mathbf{r}+\frac{\delta\mathbf{r}}%
{2},\mathbf{r}_{0},t\right)  \right)  \right]  }\ , \label{eq:fevosemicla3}%
\end{align}

\noindent As in Eq.~(\ref{eq-mAmplSemiclassico}), $\Delta S_{s,{\tilde s}}$ is
the extra contribution to the classical action that the trajectory $s$
(${\tilde s}$) acquires by effect of the perturbation $\Sigma$.

\begin{figure}[tb]
\centering \includegraphics*[width=8.3cm]{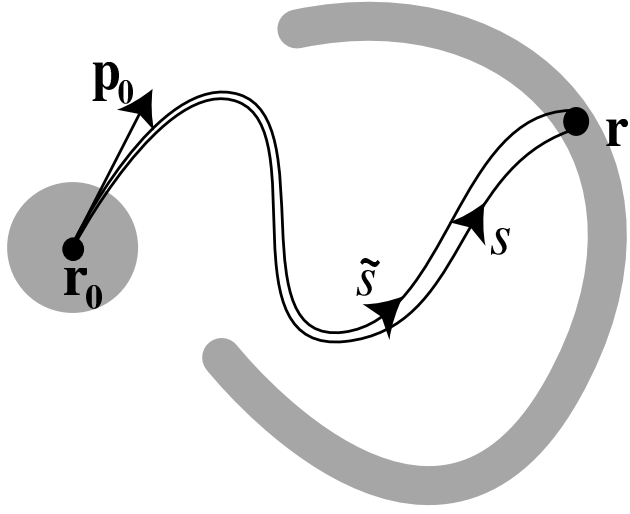}
\caption{ Classical trajectories in the manifold that evolves classically from
$\mathbf{r}_{0}$ to $\mathbf{r}$, representing the diagonal component of
$f_{\Sigma}$. The action differences $\Delta S$ of trajectories $s$ and
${\tilde s}$ are correlated. The shaded regions depict the initial and final
classical densities. }%
\label{DiagonalTraj}%
\end{figure}

We have two different cases, depending on whether or not there are
trajectories leaving from $\mathbf{r}_{0}$ with momentum close to
$\mathbf{p}_{0}$ that arrive to the neighborhood of $\mathbf{r}$ after a time
$t$. In the first case $\mathbf{r}$ is in the manifold that evolves
classically from the initial wave-packet (Fig.~\ref{DiagonalTraj}). Such a
contribution is dominated by the terms where the trajectory ${\tilde s}$
remains close to its partner $s$, and calling $f_{\Sigma}^{\mathrm{d}}$ this
diagonal component, we get

\begin{equation}
f_{\Sigma}^{\mathrm{d}}(\mathbf{r},t)=\left(  \frac{\sigma^{2}}{\pi\hbar^{2}%
}\right)  ^{d}\ \int\mathrm{d}\delta\mathbf{r}\ \sum_{s,{\tilde{s}}}%
\ C_{s}^{2}\ \exp\left[  -\frac{2\sigma^{2}}{\hbar^{2}}\left(  {\overline
{\mathbf{p}}}_{s}-\mathbf{p}_{0}\right)  ^{2}\right]  \ \exp\left[
\frac{\mathrm{i}}{\hbar}\left(  \Delta S_{s}\left(  \mathbf{r}-\frac
{\delta\mathbf{r}}{2},\mathbf{r}_{0},t\right)  -\Delta S_{{\tilde{s}}}\left(
\mathbf{r}+\frac{\delta\mathbf{r}}{2},\mathbf{r}_{0},t\right)  \right)
\right]  \ , \label{eq:fevosemicla4}%
\end{equation}

Assuming, as in Sec.~\ref{sec:semi} and Ref.~\onlinecite{cit-Jalabert-Past},
that $\mathcal{H}_{0}$ stands for a chaotic system and that the perturbation
$\Sigma$ represents a quenched disorder, upon average we obtain%

\begin{equation}
\left\langle \exp\left[  \frac{\mathrm{i}}{\hbar}\left(  \Delta S_{s}\left(
\mathbf{r}-\frac{\delta\mathbf{r}}{2},\mathbf{r}_{0},t\right)  -\Delta
S_{{\tilde{s}}}\left(  \mathbf{r}+\frac{\delta\mathbf{r}}{2},\mathbf{r}%
_{0},t\right)  \right)  \right]  \right\rangle =\exp\left[  -\frac{1}%
{2\hbar^{2}}\ A\ \delta\mathbf{r}^{2}\right]  \ , \label{eq:impaver}%
\end{equation}

\noindent where $A$ is given by Eq.~(\ref{eq:Aquencheddis}). We therefore have%

\begin{equation}
f_{\Sigma}^{\mathrm{d}}(\mathbf{r},t) = \left(  \frac{2\sigma^{4}}{\pi
\hbar^{2} A}\right)  ^{d/2} \ \sum_{s(\mathbf{r}_{0},\mathbf{r},t)}
\ C_{s}^{2} \ \exp{\left[  -\frac{2 \sigma^{2}}{\hbar^{2}} \left(
{\overline{\mathbf{p}}}_{s} - \mathbf{p}_{0}\right)  ^{2} \right]  } \ ,
\label{eq:fevosemicla5}%
\end{equation}

and the corresponding contribution to the Loschmidt echo is%

\begin{equation}
M^{\mathrm{d}}(t)=\int d\mathbf{r}\ f_{\Sigma}^{\mathrm{d}}(\mathbf{r}%
,t)=\left(  \frac{2\sigma^{4}}{\pi\hbar^{2}A}\right)  ^{d/2}\ \int
\mathrm{d}{\overline{\mathbf{p}}}\ C\exp{\left[  -\frac{2\sigma^{2}}{\hbar
^{2}}\left(  {\overline{\mathbf{p}}}-\mathbf{p}_{0}\right)  ^{2}\right]  }\ .
\label{eq:mdiag}%
\end{equation}

As in Eqs.~(\ref{eq:mwse0}) and (\ref{eq-MNDiagonal}) we have used $C$ as the
Jacobian of the transformation from $\mathbf{r}$ to ${\overline{\mathbf{p}}}$.
Now the dominant trajectories are those starting from $\mathbf{r}_{0}$ and
momentum $\mathbf{p}_{0}$. We are then back to the case of the previously
discussed (Eqs.~(\ref{eq:oversqdi3}) and (\ref{eq:mdiaglor2})) diagonal contribution.%

\begin{equation}
M^{\mathrm{d}}(t) \simeq\overline{A} \ e^{-\lambda t}, \label{eq:mdiag2}%
\end{equation}

\noindent where $C=(m/t)^{d} e^{-\lambda t}$ is assumed, and $\overline
{A}=(m\sigma/A^{1/2}t)^{d}$. The decay rate of the diagonal contribution is
set by the Lyapunov exponent $\lambda$, and therefore independent on the
perturbation $\Sigma$.

\begin{figure}[tb]
\centering \includegraphics*[width=8.3cm]{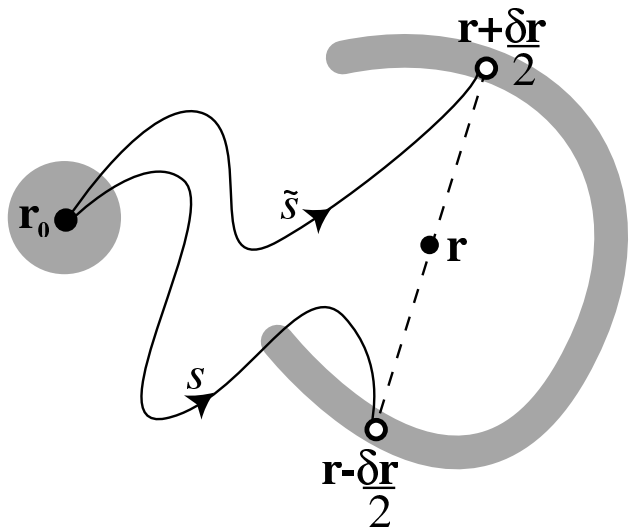}
\caption{ Non-diagonal classical contribution to the LE given by trajectories
departing from $\mathbf{r}_{0}$ and arriving to points equidistant from the
point $\mathbf{r}$ where the Wigner function is evaluated. The action
differences $\Delta S$ associated with both trajectories are uncorrelated. }%
\label{NDiagTraj}%
\end{figure}

The second possibility we have to consider is the case where there does not
exist any trajectory leaving from $\mathbf{r}_{0}$ with momentum close to
$\mathbf{p}_{0}$ that arrives to the neighborhood of $\mathbf{r}$ after a time
$t$. It is a property of the Wigner function that in the region of phase space
classically inaccessible by $X_{t}$ the points $\mathbf{r}$ half-way between
branches of the classically evolved distribution will yield the largest values
of $f_{\Sigma}$ (Fig.~\ref{NDiagTraj}). The trajectories $s$ and ${\tilde{s}}$
visit now different regions of the configuration space, therefore the impurity
average can be done independently for each of them. As in
Eq.~(\ref{eq:gauss_av2}), we have%

\begin{equation}
\left\langle \exp\left[  \frac{\mathrm{i}}{\hbar}\Delta S_{s}\right]
\right\rangle =\exp\left[  -\frac{1}{2\hbar^{2}}\left\langle \Delta S_{s}%
^{2}\right\rangle \right]  =\exp\left[  -\frac{v_{0}t}{\widetilde{2\ell}%
}\right]  _{.} \label{eq:avoftheexph2}%
\end{equation}

\noindent Such an average only depends on the length $L=v_{0}t$ of the
trajectories. Thus, after average the non-diagonal term writes%

\begin{equation}
f_{\Sigma}^{\mathrm{nd}}(\mathbf{r},t)=\left(  \frac{\sigma^{2}}{\pi\hbar^{2}%
}\right)  ^{d}\exp\left[  -\frac{v_{0}t}{\widetilde{\ell}}\right]
\ \int\mathrm{d}\delta\mathbf{r}\ \sum_{s,{\tilde{s}}}\ C_{s}C_{{\tilde{s}}%
}\ \exp{\left[  -\frac{\sigma^{2}}{\hbar^{2}}\left(  \left(  {\overline
{\mathbf{p}}}_{s}-\mathbf{p}_{0}\right)  ^{2}+\left(  {\overline{\mathbf{p}}%
}_{{\tilde{s}}}-\mathbf{p}_{0}\right)  ^{2}\right)  \right]  }\ .
\label{eq:fnondiag}%
\end{equation}

\noindent The trajectory $s$ (${\tilde s}$) goes between the points
$\mathbf{r}_{0}$ and $\mathbf{r} \mp\delta\mathbf{r}/2$. That is why the
largest values of $f_{\Sigma}^{\mathrm{nd}}(\mathbf{r},t)$ are attained when
$\mathbf{r}$ is in the middle of two branches of the classically evolved
distribution. Other points $\mathbf{r}$ result in much smaller values of
$f_{\Sigma}^{\mathrm{nd}}(\mathbf{r},t)$, since the classical trajectories
that go between $\mathbf{r}_{0}$ and $\mathbf{r} \mp\delta\mathbf{r}/2$
require initial momenta ${\overline{\mathbf{p}}}_{s}$ (${\overline{\mathbf{p}%
}}_{{\tilde s}}$) very different from $\mathbf{p}_{0}$. Thus, exponentially
suppressed contributions result.

The non-diagonal contribution to the Loschmidt echo can now be written as%

\begin{equation}
M^{\mathrm{nd}}(t)=\int\mathrm{d}\mathbf{r}\ f_{\Sigma}^{\mathrm{nd}%
}(\mathbf{r},t)=\left(  \frac{\sigma^{2}}{\pi\hbar^{2}}\right)  ^{d}%
\exp\left[  -\frac{v_{0}t}{\widetilde{\ell}}\right]  \ \left\vert
\int\mathrm{d}\mathbf{r}\ \sum_{s}\ C_{s}\ \exp{\left[  -\frac{\sigma^{2}%
}{\hbar^{2}}\left(  \left(  {\overline{\mathbf{p}}}_{s}-\mathbf{p}_{0}\right)
^{2}\right)  \right]  }\right\vert ^{2}=\exp\left[  -\frac{v_{0}t}%
{\widetilde{\ell}}\right]  \label{eq:mnd}%
\end{equation}

\noindent As in Eqs.~(\ref{eq-MNDiagonal}) and (\ref{eq-AmplitudAverage}), we
have made the change of variable from $\mathbf{r}$ to ${\overline{\mathbf{p}}%
}$, and accordingly, we have obtained the non-diagonal contribution to the LE
\cite{cit-Jalabert-Past}. As discussed before, such a contribution is a Fermi
Golden Rule like \cite{cit-Jacquod01}. In the limit of $\hbar\rightarrow0$ our
diagonal term, Eq.~(\ref{eq:mdiag2}), obtained from the final points who
follow the classical flow, dominates the LE, consistently with our findings of
Sec.~\ref{sec:univ}.

\subsection{Decoherence and emergence of classicality}

Decoherence in a quantum system arises from its interaction with an
external environment, over which the observers have no information nor control
\cite{dissipation,ZurekPT,ZurekRMP}. The states more sensitive to decoherence
are those with quantum superpositions (Schr\"{o}dinger cat states), since they
depend strongly on the information coded in the phase of the wave-function,
which is blurred by the interaction with the environment.

The studies of decoherence have traditionally considered one-dimensional
systems, and often ignored the crucial role of its underlying classical
dynamics \cite{hamburg}. On the other hand, it has been proposed
\cite{cit-Zurek-Paz} and later corroborated numerically \cite{MonteolivaPaz},
that for a classically chaotic system the entropy production rate (computed
from its reduced density matrix) is given by the Lyapunov exponent. Moreover,
as shown in Ref.~\onlinecite{cit-Jalabert-Past} and thoroughly discussed in
this work, the decay rate of the Loschmidt echo in a multidimensional
classically chaotic system becomes independent on the strength of the
perturbation that breaks the time reversal between two well-defined limits
(and set by the Lyapunov exponent). The connection between decoherence and
Loschmidt echo has been discussed in
Refs.~\onlinecite{cit-Jalabert-Past,preprintCDPZ} and has induced us to note
as $\tau_{\phi}$ the relaxation rate of the LE.

Decoherence is typically analyzed through the time decay of the
off-diagonal matrix elements of the reduced density matrix (where the
environmental degrees of freedom of the total density matrix of the system and
its environment are traced out), while the wave-function superposition
defining the LE can be cast as a trace of reduced density matrices or Wigner
functions evolving with different Hamiltonians (Eq.~\ref{eq-LEWigner}). Zurek
has recently proposed to consider the relevance of sub-Planck structure (in
phase-space) of the Wigner function for the study of quantum decoherence
\cite{cit-ZurekNature}. Considering the example of a coherent superposition of
two minimum uncertainty Gaussian wave-packets (of width $\sigma$, centered at
$\pm x_{0}$, and with vanishing mean momentum) in a one dimensional system,
where the Wigner function (up to a normalization factor) is given by%

\begin{equation}
W(x,p)=\exp\left[  -\frac{(x-x_{0})^{2}}{\sigma^{2}}-\frac{\sigma^{2}p^{2}%
}{\hbar^{2}}\right]  +\exp\left[  -\frac{(x+x_{0})^{2}}{\sigma^{2}}%
-\frac{\sigma^{2}p^{2}}{\hbar^{2}}\right]  +2\exp{\left[  -\frac{x^{2}}%
{\sigma^{2}}-\frac{\sigma^{2}p^{2}}{\hbar^{2}}\right]  }\cos\left[
\frac{2px_{0}}{\hbar}\right]  \ , \label{WignerGaussian}%
\end{equation}

\noindent it is clearly seen that in phase-space this distribution presents
two spots located around $\pm x_{0}$ positive with positive values, and
between them an oscillating structure taking large positive and negative
values (called interference fringes for their similitude with a double-slit
experience). It has then been proposed that the fringes substantially enhance
the sensitivity of the quantum state to an external perturbation. A strong
coupling with an environment suppresses the fringes, and the resulting Wigner
function becomes positive everywhere and similar to the corresponding
Liouville distribution of the equivalent classical system (with statistical
mixtures instead of superpositions) \cite{ZurekRMP}. Jacquod and collaborators
\cite{cit-JacquodSP} have contested this approach, by demonstrating that the
enhanced decay is described entirely by the classical Lyapunov exponent, and
hence insensitive to the quantum interference that leads to the sub-Planck
structures of the Wigner function.

Working with the superposition of two Wigner functions (as in the case of the
echo) and with genuinely multidimensional classically chaotic systems allows
us to give a consistent description of the connection between quantum
decoherence and the Loschmidt echo and the emergence of classical behavior.

In the previous sections, from the semiclassical evolution of the Wigner
function we were able to identify the non-diagonal component $M^{\mathrm{nd}}$
as the contribution to the LE given by the values of the Wigner function
between the branches of the classically evolved initial distribution
(Fig.~\ref{NDiagTraj}). Using the example of the two Gaussians of
Eq.~(\ref{WignerGaussian}) (but keeping in mind that the situation is more
complicated since our chaotic dynamics yields a much richer structure in
phase-space), we see that in the region between branches both of the Wigner
functions contributing to (\ref{eq-LEWigner}) are highly oscillating, and
quite different from each other. The overlap, which is perfect for zero
coupling (ensuring the unitarity requirement) is rapidly suppressed with
increasing perturbation strength. As discussed earlier in the text (see also
Refs.~\cite{cit-Jalabert-Past,cit-Jacquod01}), when $M^{\mathrm{nd}}$ is the
dominant contribution to $M$, we are in the Fermi Golden Rule regime. We have
seen that this weak perturbation regime collapses as $\hbar\rightarrow0$
(Eqs.~\ref{eq:condition} and \ref{eq-AlfaCritico}).

Beyond a critical perturbation, the diagonal component $Md$ takes over as the
dominant contribution to the LE, and is given by the values of the Wigner
function on the regions of phase space that result from the classical
evolution of the initial distribution. This is the Lyapunov regime, where the
decay rate of $M(t)$ is given by $\lambda$. Notice that this behavior is still
of quantum origin, as we are comparing the increase of the actions of nearby
trajectories by the effect of a small perturbation, assuming that the
classical dynamics is unchanged. The behavior in the Lyapunov regime does not
simply follow from the classical fidelity, where the change in the classical
trajectories is taken into account, and the finite resolution with which we
follow them plays a major role. The upper value of the perturbation strength
for observing the Lyapunov regime is a classical one, i.e. $\hbar$ independent
($\ell_{\mathrm{tr}}\simeq L$ in Sec.~\ref{subsec:leiaccs} and
Eq.~(\ref{alphap})).

For stronger perturbations (see discussions in Sects.~\ref{subsec:leiaccs} and
\ref{sec:univ}) the classical trajectories are affected and the decay rate of the
LE is again perturbation dependent. The Wigner function approach to the LE
also helps to develop our intuition about the quantum to classical transition.
The Lyapunov regime is the correct classical limit of a chaotic system weakly
coupled to an external environment.

\section{Conclusions}

\label{sec:conclusions}

In this work we have studied the decay of the Loschmidt echo in classically
chaotic systems and presented evidence for the universality of the Lyapunov
regime, where the relaxation rate becomes independent of the perturbation, and
given by the Lyapunov exponent of the classical system. Using analytical and
numerical calculations we have determined the range (in perturbation strength)
of the Lyapunov regime, its robustness respect to the classical limit, the
form of the perturbation, and the average conditions.

We presented semiclassical calculations in two different Hamiltonian systems:
a classically chaotic billiard perturbed by quenched disorder, and a Lorentz
gas where the perturbation is given by an anisotropy of the mass tensor. In
the later model, the numerical simulations were found in good agreement with
the analytical calculations, and showed that the Lyapunov decay extends
arbitrarily beyond the Ehrenfest time (where the quantum-classical
correspondence is no longer expected to hold).

Using a Wigner function representation, we have been able to present an
alternative interpretation of the two contributions to the Loschmidt echo. The
non-diagonal (Fermi Golden Rule) regime obtained for weak perturbation was
shown to arise from the destruction of coherence between non-local
superpositions thus destroying the non-classical part of the distribution. In
contrast, the diagonal (Lyapunov) regime obtained for stronger perturbation or
more classical systems was shown to be given by the classical part of the
evolved initial distribution. Thus, the Lyapunov regime is associated with the
classical evolution (even though is of quantum origin), while the Fermi Golden
Rule has a purely quantum nature. In this way, the persistence of the Lyapunov
regime after Ehrenfest time is understood as the emergence of classical
behavior due to the fast dephasing of the purely quantum terms. This is in
consistency with the understanding of the quantum-classical transition in
quantum systems coupled to an environment driven by the decoherence
\cite{ZurekPT}.

The existence and universality of an environment-independent regime and its
consequence in the phase-space behavior of the Wigner function provide a
highlight on the connection between the Loschmidt echo and quantum
decoherence. Such a connection, as well as the experiments testing the
universal behavior, are promising subjects for future research.

The universal behavior of the Loschmidt echo requires an underlying
classically chaotic system, like the ones we have considered in this work.
Hamiltonian systems with regular dynamics have been shown to exhibit an
anomalous power-law for the decay of the Loschmidt echo \cite{cit-JacquodInt}.
This behavior is quite different from the one we obtain for chaotic systems.
Therefore, we see that the Loschmidt echo constitutes a relevant concept in
the study of Quantum Chaos \cite{cit-Bohigas}. Such a connection, clearly
deserves further studies.

The Loschmidt echo in the Lorentz gas has been recently calculated for short
times \cite{GouDor}, and a rate given by twice the Lyapunov exponent has been
proposed. It would be interesting to investigate if the difference in times
scales is responsible for the departure from our results.

\acknowledgments

The authors would like to thank Ph. Jacquod, P. R. Levstein, L. F. Fo\'{a}
Torres and F. Toscano for fruitful discussions. We are grateful G.-L. Ingold and
G. Weick for their careful reading of the manuscript and valuable suggestions.
HMP is affiliated to CONICET. This work received financial support from
CONICET, ANPCyT, SeCyT-UNC, Fundaci\'{o}n Antorchas and the french-argentinian
ECOS-Sud program.

\appendix

\section{Classical dynamics with an anisotropic mass tensor}

\label{ape:reflexion}

Let us assume a particle in a free space with mass tensor $\overset
{\leftrightarrow}{m}$ surrounded by an infinite potential surface (hard wall).
Suppose that the particle departs from a point $\mathbf{r}_{0}$ at time
$t_{0}$ and arrives to a final point $\mathbf{r}$ at time $t$. We must
calculate the time $t_{\mathrm{c}}$ and position $\mathbf{r}_{\mathrm{c}}$
along the surface at which the particle collides. The action along the
trajectory is%

\begin{equation}
S=\frac{(\mathbf{r}_{c} -\mathbf{r}_{0}){\overset{\leftrightarrow}{m}%
}(\mathbf{r}_{c} -\mathbf{r}_{0})}{2(t_{c}-t_{0})}+ \frac{(\mathbf{r}
-\mathbf{r}_{c}){\overset{\leftrightarrow}{m}}(\mathbf{r} -\mathbf{r}_{c}%
)}{2(t-t_{c})} \ . \label{A-action}%
\end{equation}

We can solve the problem by minimizing the action, taking the derivative of
Eq.~(\ref{A-action}) along the surface. Introducing unitary vector
$\mathbf{n}$ normal to the surface at the point of collision, we can express
the minimization condition as%

\begin{equation}
\mathbf{n} \times\nabla_{\mathbf{r}_{c}} S = 0 \ .
\end{equation}

Denoting the initial and final velocities as $\mathbf{v}_{i}=(\mathbf{r}_{c}
-\mathbf{r}_{0})/(t_{c}-t_{0})$ and $\mathbf{v}_{f}=(\mathbf{r} -\mathbf{r}%
_{c})/(t-t_{c})$, we can write%

\begin{equation}
\mathbf{n} \times{\overset{\leftrightarrow}{m}}(\mathbf{v}_{i}-\mathbf{v}%
_{f})=0 \ .
\end{equation}

\noindent This, along with the conservation of energy $E=\mathbf{v}%
{\overset{\leftrightarrow}{m}}\mathbf{v}/2$, results in
Eqs.~(\ref{eq-ReflectionLaw}). The same result is obtained in the case of
stretched boundaries.

\section{Numerical method to simulate the quantum dynamics}

\label{ape:numerics}

In order to compute the quantum dynamics of the system we resort to a lattice
discretization (tight-binding model) in a scale $a$ much smaller that the
wavelength of the packets. This condition is relevant to accurately recover
the dispersion relation of the free particle $E_{k}=\hbar^{2}k^{2}/(2m)$, from
the energy spectrum of the (open boundaries) discretized system,
\begin{equation}
E_{k}=\frac{2\hbar^{2}}{ma^{2}}-\frac{\hbar^{2}}{ma^{2}}\left(  \cos
(k_{x}a)+\cos(k_{y}a)\right)  .
\end{equation}

\noindent Setting the lattice step as the unit length ($a=1$) we typically
worked with $R=20$, except for the calculations in Sec.~\ref{sec:univ} where,
in order to keep the precision for smaller wavelengths, $a$ was reduced
keeping the product $ka$ constant.

The discretization results in a Hamiltonian matrix whose diagonal elements are
the on-site energies. The off--diagonal elements are hopping terms
$V=\hbar^{2}/(2ma^{2})$ which is the maximum kinetic energy represented by the discretization.

The quantum dynamics on the lattice was carried out using a Trotter-Suzuki
algorithm \cite{cit-DeRaedt}, which is a remarkably precise and efficient
numerical method. At the lowest order, it is a decomposition of the evolution
operator $U$ for a small time $\tau$ in a product of analytically solvable
evolution operators. Typically one searches for a way to write the Hamiltonian
of the system as $\mathcal{H}=\sum_{k}^{Q}\mathcal{H}_{k}$, where
$\mathcal{H}_{k}$ are $2 \times2$ matrices, and thus%

\begin{equation}
U(\tau)=\exp{\left[  \mathrm{i}\mathcal{H}\tau/\hbar\right]  } \simeq
\widetilde{U}(\tau)=\prod_{k}^{Q} \exp{\left[  \mathrm{i}\mathcal{H}_{k}%
\tau/\hbar\right]  } \ ,
\end{equation}
where $U_{k}(\tau)=\exp[\mathrm{i}\mathcal{H}_{k}\tau/\hbar]$ are rotation matrices.

The highest orders involve a fractal decomposition of $\tau$ that preserves
the unitarity of the approximated evolution operator. In our calculations, a
fourth--order algorithm with a time step $\tau=0.1\hbar/V$ was precise enough
for the time regime of interest.


\begin{thebibliography}{99}
\bibitem {mistake}As noticed in Ref.~\onlinecite{cit-SmoothBilliard}, the
elastic mean-free path corresponding to a two-dimensional model of disorder
used in Refs.~\onlinecite{cit-Jalabert-Past,RapComm,JMP,tstimp,hamburg,Kbook}
should be affected by a factor of $1/2$.
\end{thebibliography}
\end{document}